\documentclass[
  journal=largetwo,
  manuscript=article-type,
  year=2025,
  volume=37,
]{cup-journal}

\usepackage{amsmath}
\usepackage[nopatch]{microtype}
\usepackage{booktabs}
\usepackage{hyperref}
\hypersetup{
    colorlinks=true,
    linkcolor=blue,
    citecolor=blue,
    filecolor=magenta,      
    urlcolor=blue,
    pdftitle={Overleaf Example},
    pdfpagemode=FullScreen,
    }
\usepackage{txfonts}

\usepackage{graphicx}
\usepackage{amssymb}
\usepackage{soul}
\usepackage{ulem}
\usepackage[]{biblatex}
\DeclareUnicodeCharacter{223C}{{}}
\def\aap{Astron. Astrophys.}            
\def\apjl{Astrophys. J.}                
\def\apjs{Astrophys. J. Suppl. Ser.}    

\title{Photometrically Selected Protocluster Candidates at z$\sim$9-10 in the JWST COSMOS-Web field}

\author{Cossas K.-W. Wu}
\affiliation{Institute of Astronomy, National Tsing Hua University, 101, Section 2. Kuang-Fu Road, Hsinchu, 30013, Taiwan}
\email[Cossas K.-W. Wu]{danniel258000@gmail.com}

\author{Chih-Teng Ling}
\affiliation{Department of Astronomical Science, SOKENDAI (The Graduate University for Advanced Studies), 2-21-1 Osawa, Mitaka, Tokyo, 181-8588, Japan}\alsoaffiliation{ National Astronomical Observatory of Japan, 2-21-1 Osawa, Mitaka, Tokyo 181-8588, Japan}

\author{Tomotsugu Goto}
\affiliation{Institute of Astronomy, National Tsing Hua University, 101, Section 2. Kuang-Fu Road, Hsinchu, 30013, Taiwan}\alsoaffiliation{Department of Physics, National Tsing Hua University, 101, Section 2. Kuang-Fu Road, Hsinchu, 30013, Taiwan}

\author{Amos Y.-A. Chen}
\affiliation{Department of Physics, National Tsing Hua University, 101, Section 2. Kuang-Fu Road, Hsinchu, 30013, Taiwan}

\author{Tetsuya Hashimoto}
\affiliation{Department of Physics, National Chung Hsing University, 145, Xingda Road, Taichung, 40227, Taiwan}

\author{Seong Jin Kim}
\affiliation{Institute of Astronomy, National Tsing Hua University, 101, Section 2. Kuang-Fu Road, Hsinchu, 30013, Taiwan}

\author{Simon C.-C. Ho}
\affiliation{Research School of Astronomy and Astrophysics, The Australian National University, Canberra, ACT 2611, Australia}\alsoaffiliation{Centre for Astrophysics and Supercomputing, Swinburne University of Technology, P.O. Box 218, Hawthorn, VIC 3122, Australia}\alsoaffiliation{OzGrav: The Australian Research Council Centre of Excellence for Gravitational Wave Discovery, Hawthorn, VIC 3122, Australia}\alsoaffiliation{ASTRO3D: ARC Centre of Excellence for All-sky Astrophysics in 3D, ACT 2611, Australia}

\author{Ece Kilerci}
\affiliation{Department of Astronomy and Space Sciences, Science Faculty, \.{I}stanbul University, Beyaz{\i}t 34119, \.Istanbul, T\"{u}rkiye}

\newcommand{\CfA}{\affiliation{Centre for Astrophysics \text{\textbar} Harvard \& Smithsonian, 60 Garden Street, Cambridge, MA 02138, USA}}
\newcommand{\JHU}{\alsoaffiliation{Centre for Astrophysical Sciences, Department of Physics and Astronomy, The Johns Hopkins University, 3400 N Charles St., Baltimore, MD 21218, USA}}
\newcommand{\STScI}{\alsoaffiliation{Space Telescope Science Institute (STScI), 3700 San Martin Drive, Baltimore, MD 21218, USA}}
\author{Tiger Yu-Yang Hsiao} 
\CfA \JHU \STScI

\author{Yuri Uno}
\affiliation{Department of Physics, National Chung Hsing University, 145, Xingda Road, Taichung, 40227, Taiwan}

\author{Terry Long Phan}
\affiliation{Institute of Astronomy, National Tsing Hua University, 101, Section 2. Kuang-Fu Road, Hsinchu, 30013, Taiwan}


\begin{document}

\begin{abstract}
High-redshift protoclusters are crucial for understanding the formation of galaxy clusters and the evolution of galaxies in dense environments. The James Webb Space Telescope (JWST), with its unprecedented near-infrared sensitivity, enables the first exploration of protoclusters beyond \( z > 10 \). Among JWST surveys, COSMOS-Web Data Release 0.5 offers the largest area (\(\sim 0.27\) deg\(^2\)), making it an optimal field for protocluster searches. In this study, we searched for protoclusters at \( z \sim 9-10 \) using 366 F115W dropout galaxies. We evaluated the reliability of our photometric redshift by validation tests with the JADES DR3 spectroscopic sample, obtaining the likelihood of falsely identifying interlopers as $\sim25\%$. Overdensities (\(\delta\)) are computed by weighting galaxy positions with their photometric redshift probability density functions (PDF), using a 2.5 cMpc aperture and a redshift slice of \(\pm 0.5\). We selected the most promising core galaxies of protocluster candidate galaxies with an overdensity greater than the 95th percentile of the distribution of 366 F115W dropout galaxies. The member galaxies are then linked within an angular separation of 7.5 cMpc to the core galaxies, finding seven protocluster candidates. These seven protocluster candidates have inferred halo masses of \( M_{\text{halo}} \sim 10^{11} M_{\odot} \). The detection of such overdensities at these redshifts provides a critical test for current cosmological simulations. However, confirming these candidates and distinguishing them from low-redshift dusty star-forming galaxies or Balmer-break galaxies will require follow-up near-infrared spectroscopic observations.
\end{abstract}

\section{Introduction}
\label{Intro.}
The subtle density fluctuations in the early Universe lead to the aggregation of galaxies.
Over time, the galaxies formed protoclusters, which are the first assemblies of galaxies in which all the haloes/galaxies will eventually merge into the final galaxy cluster in the local Universe \citep{2015Muldrew}.
As the most extreme matter overdensities within the cosmic web, galaxy clusters reside at the nodes of this large-scale structure permitted by the standard cosmological framework of hierarchical structure formation \citep{1978White_Rees}, protoclusters are also expected to trace the characteristics of the large-scale structures (LSS).
Given their rarity, particularly during the early Epoch of Reionisation (EoR), confirming the existence of protoclusters not only serves as an observational milestone but also provides crucial insights into the evolutionary history of the Universe.
Despite their significance, our understanding of how galaxies evolve and organise into protoclusters, eventually becoming part of the LSS, remains limited. This knowledge gap arises primarily due to the challenges in achieving the deep sensitivity required at near-infrared wavelengths, where much of the critical data about these distant and early structures resides. Addressing these challenges is essential to unravel the complexities of galaxy evolution and the emergence of large-scale cosmic structures.

During the past half-decade, astronomers have found many galaxy protoclusters at $2 < z < 8.5$ \citep[][etc.]{2019Higuchi, 2019Harikane, 2021Hu, 2021Polletta, Laporte2022, 2024Brinch, Helton2024a, 2024Wang, Chen2024, 2025Fudamoto}.
We expect to extend our knowledge about the evolution of galaxies in the distant Universe ($z > 8$), by utilising the James Webb Space Telescope (\cite{2023Gardner}, JWST) to reveal the mysterious early galaxies with its unprecedented sensitivity and resolution in Infrared (IR).

Recent studies from JWST have begun to uncover the existence of spectroscopically confirmed high-z protoclusters:
\citet{Laporte2022} found a lensed protocluster candidate at $z = 7.66$ behind the background of the SMACS0723-7327 galaxy cluster, and \citet{Chen2024} reports an overdensity of Lyman-$\alpha$ galaxies at $z = 7.88$ in the CEERS EGS field.
Above $z > 8$, \citet{Larson2022} discovered a potential protocluster comprising a pair of galaxies at $z_{\text{spec}} \sim 8.7$ with a separation of 3.5 proper Mpc ($\sim$ 34 comoving Mpc (cMpc)) among 11 galaxies at $z > 8$ in the CANDELS survey fields ($\sim 800$ arcmin$^{2}$). 
Similarly, \citet{Helton2024b} confirmed an overdensity includes six galaxies at $z_{\text{spec}}=8.22$ spanning across 2.2 cMpc.

Observed quantities such as the number density of protoclusters show us the overview of the structural development in the early Universe.
However, the previous JWST vision from \citet{Helton2024b} is limited to the survey area of 134 arcmin$^{2}$ of JWST Advanced Deep Extragalactic Survey (JADES. \cite{2023Eisenstein}).
A larger survey area will put more constraints on the number density of the protoclusters.
The COSMOS-Web survey from Cycle 1 JWST treasury program~\citep{2023Casey} provides an opportunity to search for high-redshift protoclusters with extensive coverage and multi-wavelength auxiliary data.
The survey area is superior to others because it has been observed by major space-based (HST, Spitzer, Herschel, and others) and ground-based telescopes (Keck, Subaru, and others).
Also, the COSMOS-Web field has the largest survey area ($\sim$0.5 deg$^2$) among the JWST surveys to date.
Furthermore, simply scaling the number density of very distant protoclusters (candidates) found by \cite{Larson2022} and \cite{Helton2024b}, we expected that the field of view of COSMOS-Web would contain $\sim 1-2$ protoclusters in $z\sim9$ using the survey area ratio.
Based on the reasons above, COSMOS-Web naturally becomes the best site for discovering any potential protoclusters in the early Universe.

Here is an overview of this work.
In section \ref{Data}, we introduce the basic data properties of the COSMOS-Web field. 
We summarise the construction of a multi-band catalogue in \ref{Mul-wave. Phot. Cat.}. 
We then explain the methodology in section \ref{Methodology}, including applying colour-selection criteria to the photometry results. We discuss the results of spectral energy distribution(SED) fitting to high-redshift galaxy candidates in section \ref{SED&Photo-z}. 
Lastly, in section \ref{Results}, we consider the possible clustering of the early galaxies by the overdensity.

If else specified, we adopt the {\it Planck15} cosmology \citep{Planck2016}, i.e., $\Lambda$ cold dark matter cosmology with ($\Omega_{m}$, $\Omega_{\Lambda}$, $\Omega_{b}$, $h$)=(0.307, 0.693, 0.0486, 0.677).

\section{Data}
\label{Data} 
This work utilizes the public data release 0.5 (hereafter DR0.5) from the COSMOS-Web survey \citep[GO \#1727, PIs Kartaltepe \&\ Casey;][]{2023Casey}.
The DR0.5 contains the observations undergone during early January and June / July 2023 (observation Nos. 043-048 and 078-153), and covers 0.29 deg$^{2}$ in total.
This release separates the data into 10 mosaic images, A1-A10, accordingly. 
Near-Infrared Camera (NIRCam) observations are conducted with four filters (F115W, F150W, F277W, F444W), while Mid-Infrared Instrument (MIRI) observations are conducted with a single filter (F770W).
The data is available on the official website from the COSMOS team:
\footnote{\url{https://cosmos.astro.caltech.edu/page/cosmosweb-dr}}.

    \subsection{Data Reduction}
    \label{Data Reduction}
    The raw images from \textit{JWST} were processed using the JWST Calibration Pipeline v1.10.0 with CRDS Context pmap = 1075 \citep{2022Bushouse}.
    The COSMOS team implemented extra customised modifications and reduction techniques to enhance the data product.
    The astrometric precision of the NIRCam mosaics has been improved through alignment with the Gaia-DR3 \citep{2022Weaver} and the COSMOS2020 catalogue. This was enabled by using a reference catalogue based on COSMOS HST/F814W imaging data, which had been meticulously reprocessed following the methodology originally described in \citet{2011Koekemoer}.
    
    \subsection{Multi-wavelength Photometric catalogue}
    \label{Mul-wave. Phot. Cat.}
    In this section, we describe the details of building our multi-wavelength photometric catalogue.
    We use \textit{Source-Extractor} V2.19.5 \citep[hereafter SEx,][]{1996Bertin} for source extraction and photometry.
    We use the default parameter except for ${\tt DETECT\_THRESH}=2.5$, ${\tt DEBLEND\_NTHRESH}=48$, and ${\tt MAG\_ZEROPOINT}$ in the configure file for SEx.
    ${\tt MAG\_ZEROPOINT}$ for each filter is set to the numbers obtained by the first equation on the NIRCam zeropoints website \footnote{\url{https://jwst-docs.stsci.edu/jwst-near-infrared-camera/nircam-performance/nircam-absolute-flux-calibration-and-zeropoints##gsc.tab=0}}.
    To efficiently identify the dropout sources by colours, we enable the dual mode in the SEx. 
    Dual mode allows us to perform force photometry on measurement images for every source that is detected in the detection image, but it would require identical pixel coordinates between both images.
    For each mosaic image from DR0.5, we co-add F150W, F277W, and F444W as the detection image with the default parameter in the \textit{SWarp} \citep{Bertin2010}.
    Measurement images are simply the re-projected and resampled mosaic images from DR0.5 on the same x,y coordinates as the detection images. 
    Both the detection image and measurement image are resampled to the pixel scale of 0.03 arcseconds by the authors as part of a customised preprocessing step.
    
    We measure the flux within a 0".15 radius aperture, which is slightly larger than the FWHM of the point-spread function in the F444W band (0".145), then apply the aperture correction with the interpolated coefficients derived from the JWST CRDS\footnote{\url{https://jwst-crds.stsci.edu/browse/jwst_nircam_apcorr_0004.fits}}.

    Our JWST photometric catalogue is cross-matched to the COSMOS-2020 catalogue \citep{2022Weaver} with a maximum error tolerance of 75 milliarcseconds (2$\sigma$) for each pair.
    We use the median of the difference in each coordinate to align HST and JWST cutout images.
    The COSMOS-2020 catalogue \citep{2022Weaver} concludes the previous results from various telescopes, such as UVISTA, HSC, and HST. 
    We further compare the observed AUTO magnitude between JWST NIRCam F150W and the UVISTA H band for every matched source.
    We found that the median of the difference between the UVISTA H band AUTO magnitudes and JWST NIRCam F150W is less than $< 25\%$ down to $\sim$27.0 AB magnitude.

    After cross-matching the COSMOS-2020 catalogue, we performed forced photometry on the HST image with \textit{Photoutils} for the unmatched sources.
    To evaluate the flux uncertainty of forced photometry on the F814W band, we placed a 0.3"-0.6" annulus centred on the source to determine the flux uncertainty with \textit{Photutils}. 
    The flux uncertainty in the other JWST bands was measured with the SEx using the same 0.3"-0.6" annulus that included photon and detector noise.
    

\section{Methodology}
\label{Methodology}
In this section, we first follow the colour criteria outlined by \citet{2023Harikane} to identify F115W dropout galaxies. Then, we estimate the photometric redshift (photo-z) and generate a redshift probability distribution function (PDF) for the selected candidates. Finally, we explain the method for estimating the number density of the galaxies as well as their overdensity.

    \subsection{colour selection}
    \label{colour selection}
    \citet[][H+23]{2023Harikane} set an example for selecting high-redshift galaxy candidates by simple yet efficient colour selection in recent JWST observations.
    The available four major NIRCam filters in COSMOS-Web (F115W, F150W, F277W, and F444W) could only have identical selection criteria (Eq. \ref{colour-criteria}) for picking up F115W dropouts in H+23. 
    
    \begin{align}
    \label{colour-criteria}
        F115W-F150W &> 1.0 \ \wedge \nonumber \\
        F150W-F277W &< 1.0 \ \wedge \\
        F115W-F150W &> F150W-F277W + 1.0 \nonumber
    \end{align}
    
\begin{figure}
        \centering
        \includegraphics[width=\columnwidth]{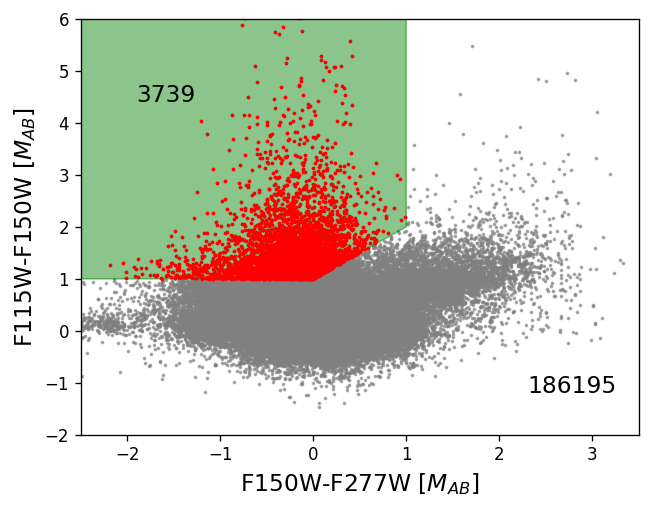}
        \caption{The two-colour diagram for the sources have $S/N > 2$ in F115W+F277W+F444W detection image of COSMOS-Web DR0.5 \cite{2023Casey}. The green area shows colours that satisfy the F115W-dropout criteria from \cite{2023Harikane}. The colour is measured with a 0.3" diameter circular aperture. The numbers indicate how many sources there are in each subset.}
        \label{C.-C. Diagram}
    \end{figure}

    We measure the colours of each object using a fixed circular aperture with a diameter of 0.3 arcseconds and apply the aperture correction from the coefficients listed in the JWST CRDS.
    Figure \ref{C.-C. Diagram} shows the distribution of our selected galaxies on the two-colour diagram.
    In total, we selected 3739 objects with matched colours.
    
    For sufficient robustness of the selection of dropout galaxies, we adopt a 2$\sigma$ flux density cut in the F814W and F115W bands within a circular aperture of 0.3" diameter \citep{2022Harikane, 2023Harikane, Hainline2024}. We further rejected 2864 objects out of 3739 objects with S/N > 2 in any of the F814W and F115W filters based on the results of forced photometry to avoid interference, leaving us with 875 objects after this process.
        
    \subsubsection{Comparison with JADES}
    \label{JADES_test}
\begin{figure}
            \centering
            \includegraphics[width=\columnwidth]{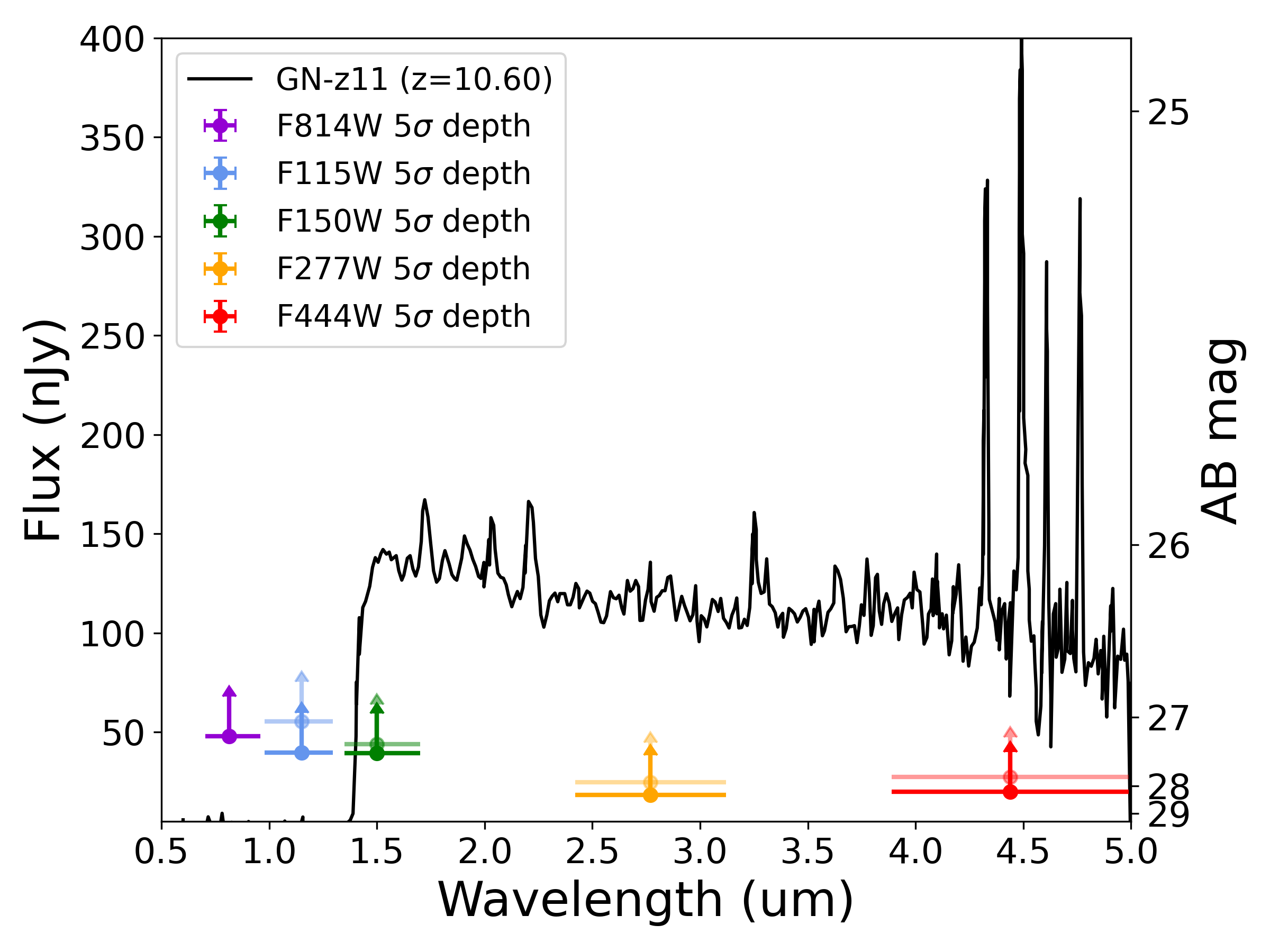}
            \caption{Spectral energy distribution (SED) of the spectroscopically confirmed galaxy GN-z11 at redshift z=10.60 (black line), overlaid with the 5$\sigma$ detection limits of various filters used in this study. The coloured upward arrows indicate 5$\sigma$ limiting depths in each band: F814W (purple), F115W (light blue), F150W (green), F277W (orange), and F444W (red). For each band, the fainter (transparent) arrows denote the shallower 5$\sigma$ depths reached in approximately 50\% of the survey area, due to non-uniform coverage and exposure time. Fluxes are shown in nJy on the left y-axis, with the corresponding AB magnitudes on the right y-axis. Horizontal error bars represent the approximate width of each filter’s transmission curve. This figure illustrates the ability of the JWST NIRCam bands to probe the rest-frame UV-to-optical emission of galaxies at z>10.}
            \label{Depth_Comparison}
        \end{figure}

        Moreover, to show the reliability of the colour-selected samples, we compare them with spectroscopically confirmed galaxies. 
        We first prepared all sources that have both NIRCam and NIRSpec observations in the JADES field. 
        According to JADES DR3 \citep{Francesco2025}, we then selected galaxies only with the quality of the spectra ranks A, B, and C, which have secure and visually identified emission lines. 
        The flux uncertainty of this photometric data set is then downgraded to the typical noise levels expected in COSMOS-Web. 
        In Figure \ref{Depth_Comparison}, we show the spectra of GN-z11 with the depth of the COSMOS-Web at each wavelength, indicating that the depth is enough to detect bright galaxies up to z$\sim$11.
        However, we found that no galaxies in the JADES DR3 catalogue satisfy both the COSMOS-Web colour criteria and the signal-to-noise thresholds while also being brighter than the COSMOS-Web limiting magnitudes in the F150W, F277W, and F444W bands.
        We attribute the lack of bright sources to the small field of view of the JADES.
        Therefore, to further test the reliability of the colour-selected sample, we relaxed the signal-to-noise constraints and applied only the colour criteria.
        
\begin{figure}
            \centering
            \includegraphics[width=\columnwidth]{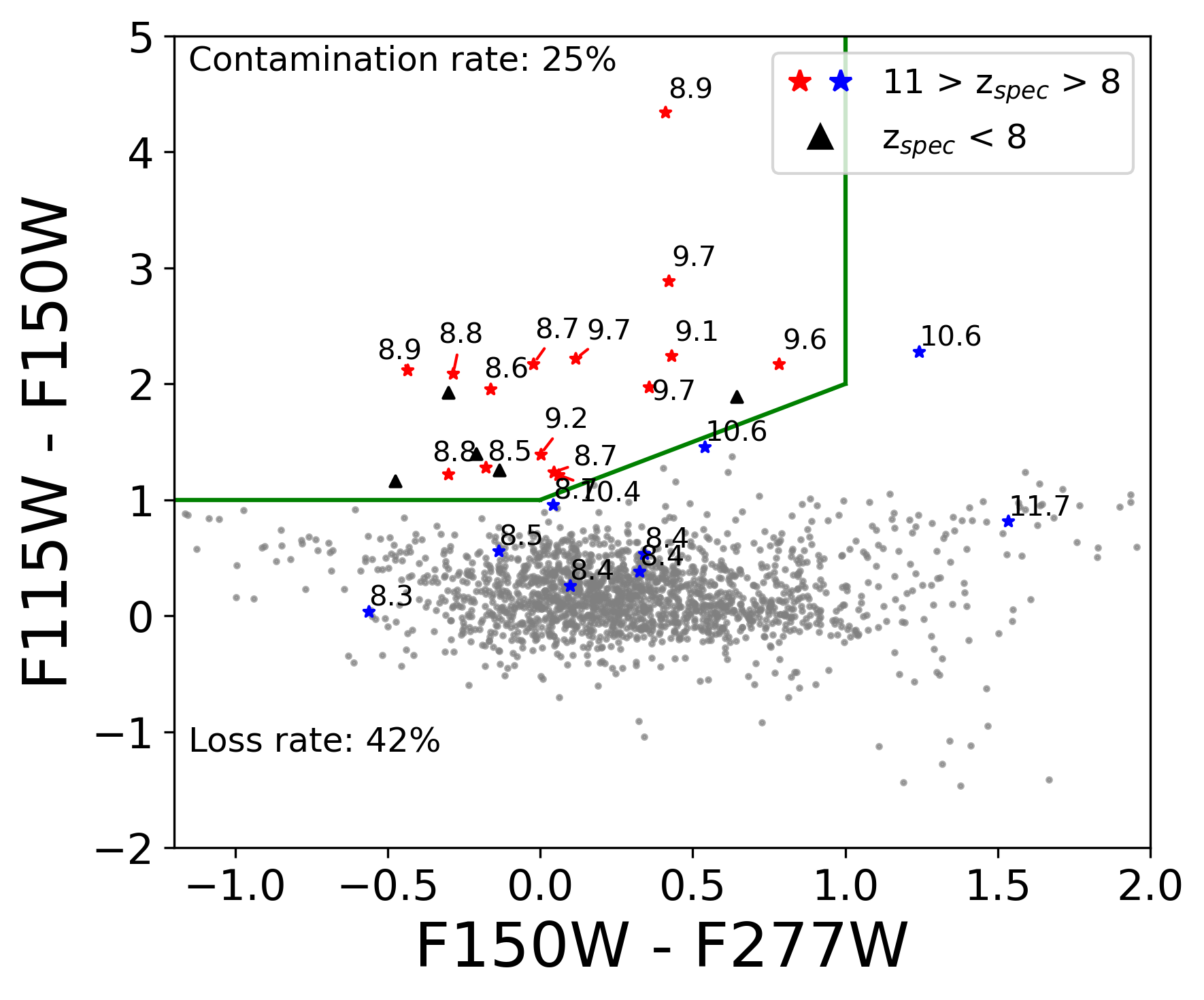}
            \caption{The two-colour diagram for sources from the JADES DR3 matching COSMOS-Web limiting magnitudes.
            Scattered grey dots represent all objects that have both NIRCam and NIRSpec data. The green polygon delineates our F115W-dropout selection criteria. coloured stars are spectroscopically confirmed galaxies with $8.0\leq z_{\text{spec}}\leq12.0$; the over-plotted numbers mark their spectroscopic redshifts. Red/Blue colours mark galaxies that satisfy/fail the colour criteria. Black triangles indicate $z_{\text{spec}}<8.0$ that satisfy the colour criteria sources. The text annotations quote a contamination rate of 25\% (ratio between galaxies with $z_{\text{spec}}<8.0$ that satisfy the colour criteria (N=5) to the number of the source satisfy the colour criteria (N=20)) and a loss rate of 42\% (fraction of $8.0\leq z_{\text{spec}}\leq12.0$ galaxies that do not satisfy the colour criteria (N=11) to the number of all $8.0\leq z_{\text{spec}}\leq12.0$ galaxies (N=26).}
            \label{JADES Comparison}
        \end{figure}

        Figure \ref{JADES Comparison} shows the contamination rate in the selection of the colour criteria, which is the fraction of 8 < z < 12 candidates that are, in fact, lower-redshift galaxies, is 25\%. The loss rate, the fraction of 8 < z < 12 candidates that do not satisfy the colour criteria, is $\sim$42\%.
        
        Also, at redshifts $z\geq11$, the effect of the intergalactic medium (IGM) on observed F150W fluxes becomes increasingly severe. In particular, the damping wing of Lyman-$\alpha$ can extend well into the F150W filter, significantly suppressing the observed continuum flux beyond the nominal wavelength of the Lyman-$\alpha$ break. This results in diminished detectability for sources at $z\geq11$. Consequently, our colour selection technique becomes less efficient at the highest redshifts, as the $z\geq11$ galaxy exhibits a F150W-F277W colour far beyond in the Figure \ref{JADES Comparison}. Therefore, it is likely that some extremely high-redshift galaxies are missed due to IGM attenuation.

    \subsection{SED Fitting and Photometric Redshifts}
    \label{SED&Photo-z}
        Here we perform SED analysis mainly with \textit{CIGALE} \citep{Boquien2019} v2022.1, and test with \textit{EAZY} \citep{Brammer2008} as a comparison to the derived redshifts and galaxy properties.
        We integrate all the available photometric data from the \textit{JWST} and \textit{HST} in the COSMOS-Web field.
        This includes F814W, F115W, F150W, F277W, and F444W.
        Following the analysis in Section \ref{JADES_test}, we further compare the photometric redshift performances of these two codes using galaxies with spectroscopic redshifts in the JADES DR3 catalogue. We only used filters available in the COSMOS-Web and specifically downgraded JADES photometry to COSMOS-web depths as in Sections \ref{SED_CIGALE} and \ref{SED_EAZY}.

        \subsubsection{SED with \textit{CIGALE}}
        \label{SED_CIGALE}
            \textit{CIGALE} provides comprehensive modelling that allows us to more carefully evaluate how different physical assumptions (nebula emission, dust, etc.) might affect the inferred high-redshift solutions.
            We adopt the \texttt{dustatt\_modified\_CF00} module for dust attenuation, single stellar population (SSP, \citet{2003Bruzual}), and \texttt{sfhdelayed} module for star-forming history. Among the settings in the \textit{CIGALE}, the photo-z are fitted from 0 to 12 with an increment of 0.06.
        
            The strong emission-line galaxies at $z<6$ can be concerning interlopers that masquerade as $z\sim16$ Lyman-break galaxies (Arrabal Haro et al. 2023). In the sense that a potentially significant contribution from H$\beta$+OIII to the F277W flux, and H$\alpha$ to the F444W flux for emission galaxies at $z\sim4-5$. Therefore, including nebula emission in the CIGALE modelling should result in a better chance of distinguishing the interlopers.
            
            To even shorten the time for calculating the best-fit model in \textit{CIGALE}, we did not include AGN models, \textit{X-ray}, and \textit{Radio} parameters.

\begin{figure}
            \centering
            \includegraphics[width=\columnwidth]{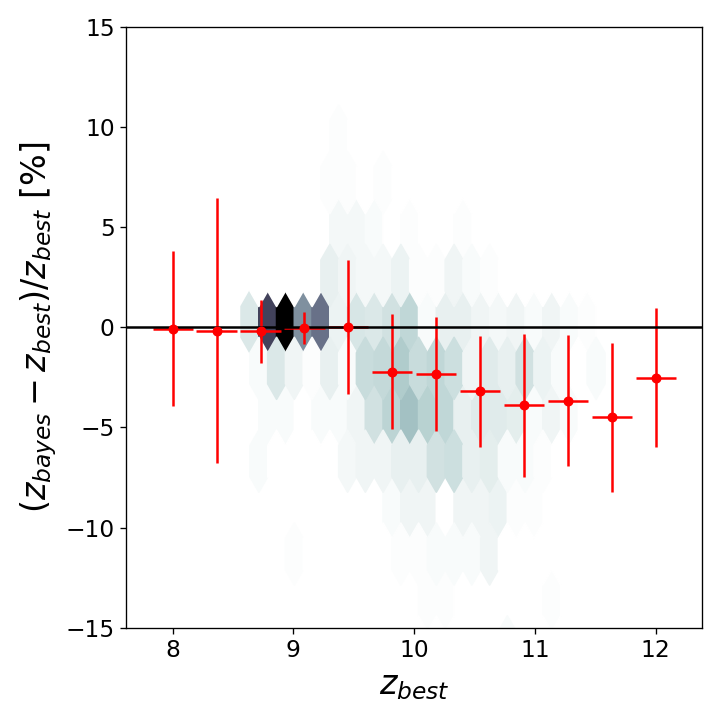}
            \caption{The relative percentage deviation of photo-z derived from CIGALE for the best-fit model and the Bayesian estimation. The colourmap indicates the number of sources in each hexagonal bin. Red scatter points represent the median of certain redshifts and the corresponding standard deviation. The Bayesian estimation does not deviate $> 7.5\%$ from the best-fit model across 8 $\leq z \leq$ 12 for most sources.}
            \label{Error Ratio}
        \end{figure}

            The photometric results of \textit{CIGALE} come in two parts, with the best fit and Bayesian. 
            The ratio between best-z and Bayesian-z of all of the remaining galaxy samples is shown in Figure \ref{Error Ratio}.
            Since \textit{CIGALE} does not calculate the error of the best-z, we adopt the 16 and 84-percentile of the PDF to present its lower-/upper error.
        
        \subsubsection{SED with \textit{EAZY}}
        \label{SED_EAZY}
            \textit{EAZY} combines user-supplied templates to derive observed photometry for a given galaxy. The template set we used in \textit{EAZY} includes the \texttt{tweak\_fsps\_QSF\_12\_v3} set of 12 flexible stellar population synthesis
            \citep[FSPS;][]{Conroy2010} templates recommended by the \textit{EAZY} documentation. We also included the nine templates (Set1+3+4) designed for high-redshift galaxies from \citet{Larson2022}, and seven JADES SED templates used by \citet{Hainline2024}.
            
            To prevent the F814W flux upper limit from overly constraining the fits and to account for any photometric calibration uncertainties during the reprojecting, we set \textit{EAZY} to have an additional photometry uncertainty of 10\%. We did not adopt any apparent magnitude priors in the \textit{EAZY} parameter to avoid dropping the faint sources.
        
\begin{figure}
            \centering
            \includegraphics[width=\columnwidth]{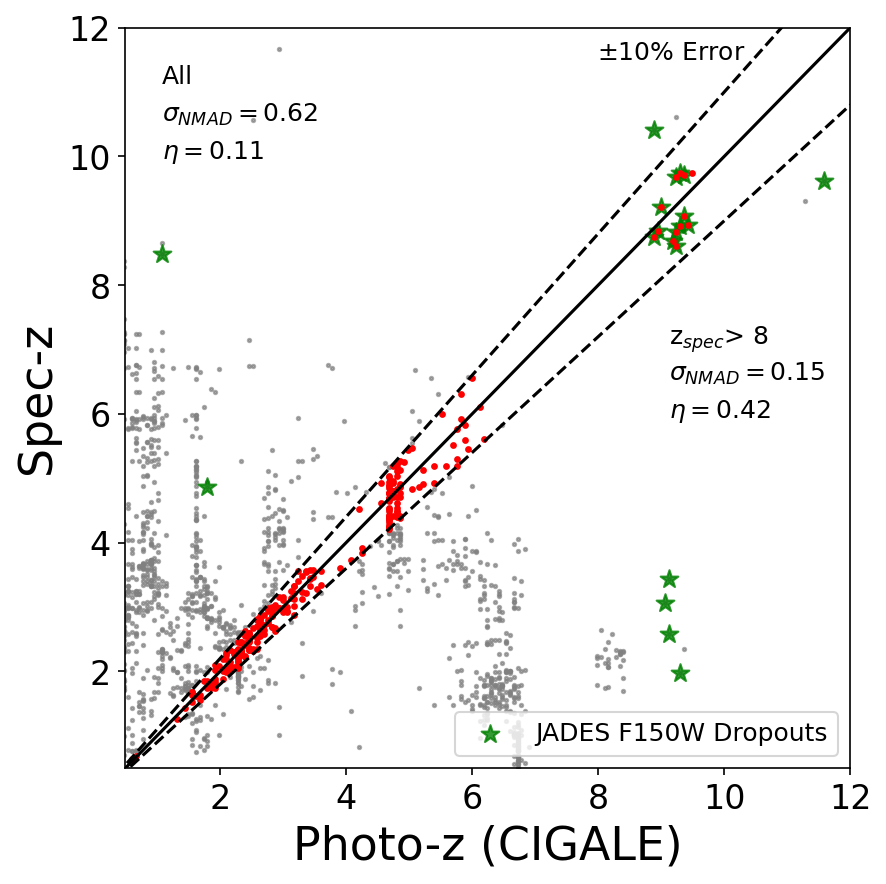}
            \includegraphics[width=\columnwidth]{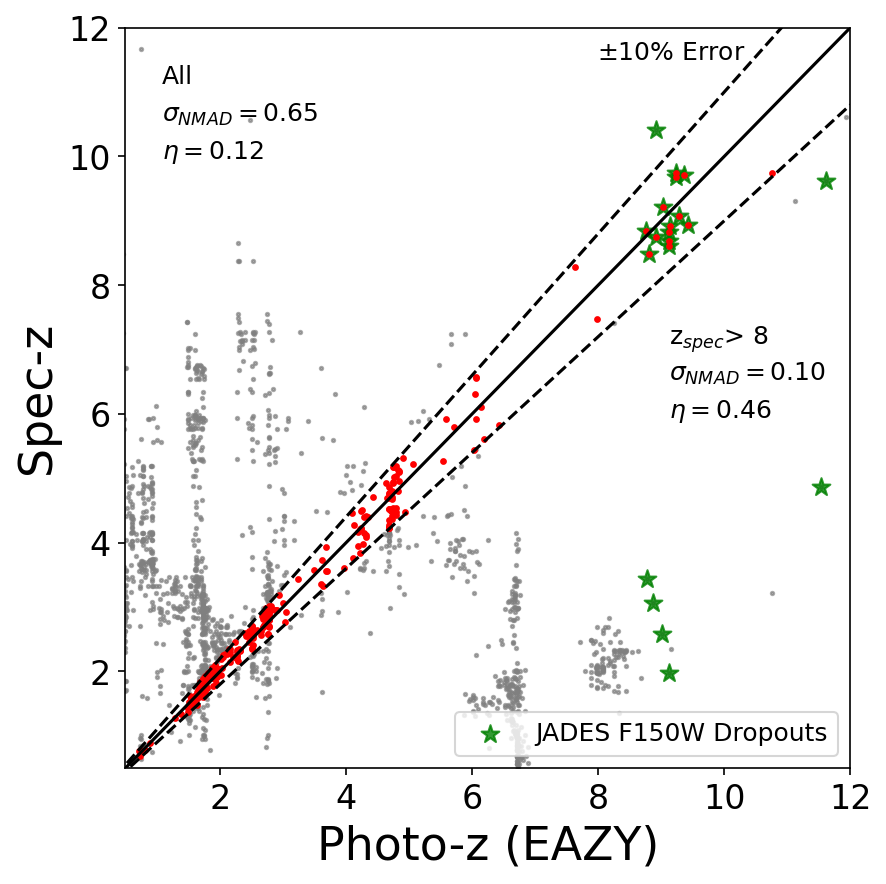}
            \caption{Comparison between photometric and spectroscopic redshifts from \textit{CIGALE}/\textit{EAZY} and spectroscopic redshifts from JADES DR3. Grey points represent all galaxies with both photometry and spectroscopic redshifts in the JADES field.
            Note that we did not use filters not available in the COSMOS-Web, and the JADES photometry is downgraded to the COSMOS-Web quality. Green stars denote JADES galaxies that satisfy our F150W-dropout colour criteria. The solid black line indicates the one-to-one correspondence ($z_{\mathrm{spec}} = z_{\mathrm{phot}}$), and the red points highlight sources within the dashed lines of 10\% deviation. 
            For the high-redshift subset ($z_{\mathrm{spec}} > 8$), we find a significantly improved normalised median absolute deviation (NMAD) and outlier fraction. This figure demonstrates the performance of photometric redshift estimation under COSMOS-Web-like conditions.}
            \label{SED Comparison}
        \end{figure}

        In Figure \ref{SED Comparison}, the resulting statistics included an estimate of the normalised median absolute deviation (NMAD), $\sigma_{\text{NMAD}}$, and the 10\% outlier rate, $\eta$:
            
        \begin{equation}
        \label{sigma_NMAD}
            \sigma_{\text{NMAD}} = 1.48 \times \text{median}(|\Delta z - \text{median}(\Delta z)|)
        \end{equation}
        \begin{equation}
        \label{eta}
            \eta = \frac{N_{|\Delta z|>0.1z}}{N_{\text{tot}}}
        \end{equation}
        
        As Figure \ref{SED Comparison} shows, both software yield similar results in terms of $\eta$ and $\sigma_{NMAD}$, which equals 0.50(0.58) and 0.15(0.10), from \textit{CIGALE}(\textit{EAZY}). Considering that \textit{CIGALE} could provide a simultaneous fitting of the physical parameters for the analysis, we use \textit{CIGALE} throughout the manuscript.
        
\begin{figure*}
            \centering
            \includegraphics[width=\columnwidth]{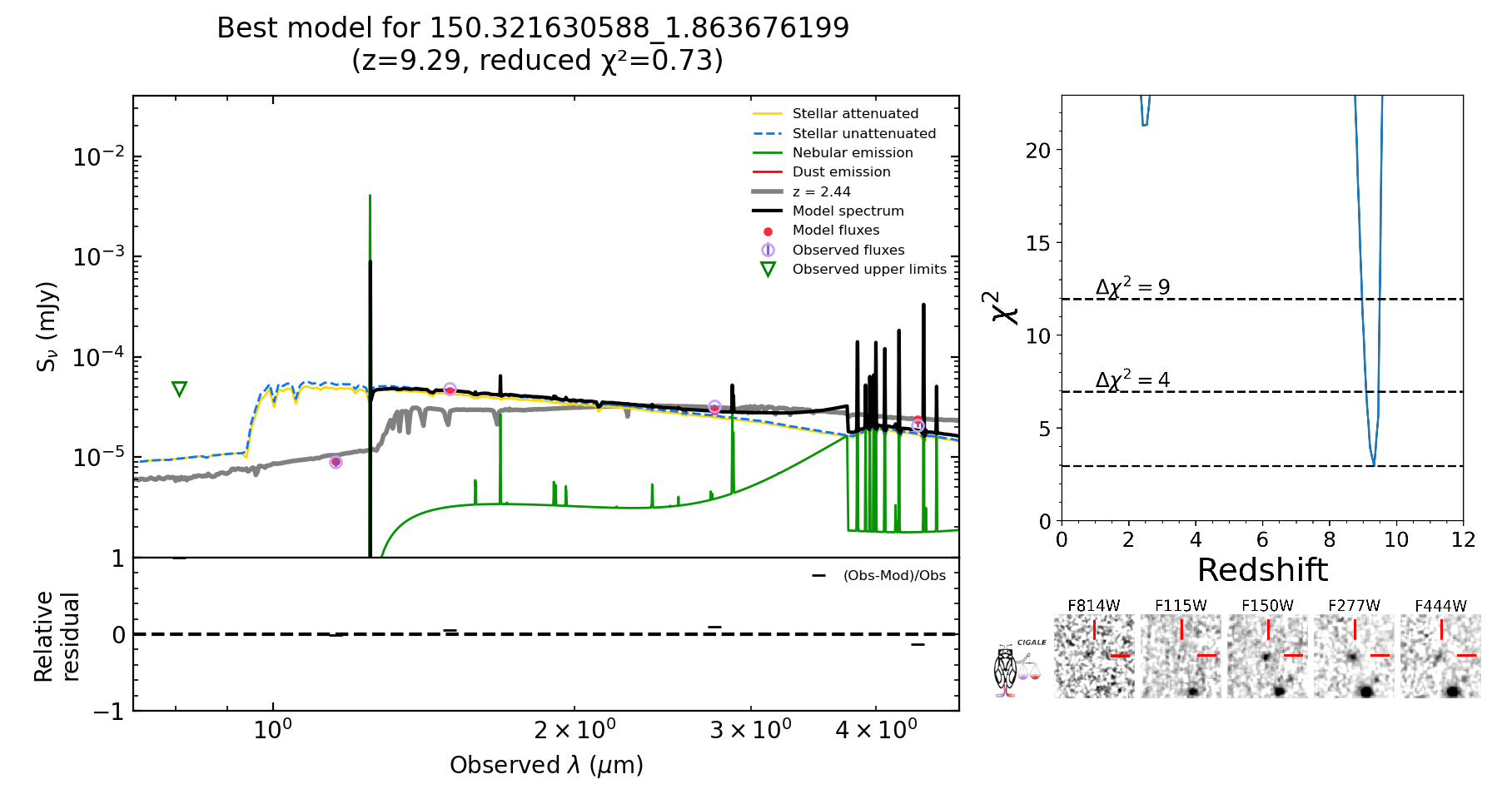}
            \caption{CIGALE best-fit spectral energy distribution and redshift likelihood for galaxy.
            \textbf{Left}: The black curve shows the total model SED corresponding to the minimum-$\chi^2$ z > 8 solution. Coloured components indicate the attenuated stellar continuum (yellow), grey intrinsic stellar emission (blue dashed), nebular lines and continuum (green), and thermal dust emission (red). Magenta circles mark the observed NIRCam/IRAC fluxes, while the green triangle denotes a 2-sigma upper limit. We added the best $z_\mathrm{phot} < 8$ solution as a grey line for reference.
            \textbf{Top-right}: variation of $\chi^2$ with redshift. The sharp minimum at $z\approx9.3$ and the absence of significant secondary solutions ($\Delta\chi^2 \leq 9$) at lower redshift (z<8) confirm the robustness of the high-z interpretation.
            \textbf{Bottom-right}: 2" × 2" cutouts in HST/ACS F814W and JWST/NIRCam F115W, F150W, F277W, and F444W (left to right). Red tick marks (0.5" in length) identify the target. The non-detections in F814W and F115W, coupled with clear detections long-ward of 1.5 $\mu$m, are consistent with a Lyman-break galaxy at $z\approx9.3$.}
            \label{SED Example}
        \end{figure*}

    To further demonstrate the robustness of photometric redshift from \textit{CIGALE}, we followed \citet{2023Harikane}, \citet{Finkelstein2023}, and \citet{Hainline2024}, referring to the criteria for selecting a high-z solution is much preferred over the low-z solution by $\Delta\chi^2 = \chi^{2}_{(z<8)} - \chi^{2}_{(z>8)} >= 9$. 
    Figure \ref{SED Example} shows the SED result from \textit{CIGALE} with the corresponding $\chi^2$ distribution for passed/failed galaxies.
    
    The $\chi^{2}$ distribution is converted from the PDF of redshift from \textit{CIGALE} with the same equation implemented in \textit{EAZY} (Eq. \ref{Delta_chi2}).
    \begin{equation}
    \label{Delta_chi2}
        P(z) =e^{(\chi^{2} - \chi^{2}_{min})/2}
    \end{equation}

    We also selected sources that have the best photo-z solution occurring at z>8. In total, this leaves us with a final sample of 366 sources.

    \subsection{Clustering of the Galaxies}
    \label{overdensity}
        \subsubsection{Weighted Number Density}
        \label{weighted Num.-Den.}
        Since the colours of the Balmer-break galaxies at a lower-z ($z\sim2-3$) and Lyman-$\alpha$-break galaxies (at $z\sim9$) have degeneracies, our galaxy sample could easily be contaminated by the low-z interlopers.
        To avoid confusion, we use the Bayesian analysis in \textit{CIGALE} to retrieve the PDF of redshift to weigh the number density of galaxies at higher redshift.
        
        In the total of 366 F115W-dropouts, we only found 10 galaxies that with $z_{\text{best}}\geq 11$, and 13 galaxies with $z_{\text{best}}\leq 9$, which are statistically insufficient to evaluate the number-/overdensity at has $z_{\text{best}}\leq 9$ and $z_{\text{best}}\geq 11$. As a result, we probe the clustering of 343 galaxies to those located only between $9.0\leq z_{\text{best}} \leq11$.

        Since the median of the resulting error varies and is non-negligible ($\pm\Delta z \sim$ 0.5), we decided to measure the local number density centred at each galaxy, integrating the PDFs of every galaxy within an aperture with a certain radius and weighing them correspondingly (Eq. \ref{Eq. Weights}).
        The detailed reasoning for the chosen radius is in the next section (Sec. \ref{Sigma & Radius}).
        The redshift interval in the equation is centred on the best-fit redshift of the galaxy; we want to compute its local number density with the $\pm\Delta z=0.5$ redshift interval, which combined with the aperture, makes the local volume a cylinder.
        The length of the cylinder ($\pm\Delta z\sim$ 0.5) is equivalent to a distance of $\sim 250$ cMpc at $z=9.0$.
        \begin{equation}
            \label{Eq. Weights}
            W = \int_{z_{1}}^{z_{2}} P(z) \,dz
        \end{equation}

        At the edge of the survey field, we also scale the number density by the fraction of the volume of the cylinder within the survey field.
        To avoid overestimating the number density at the edge of the survey area, we discarded the measurements that had a scale factor greater than 2.
        
        The overdensity ($\delta$) is calculated as Eq. \ref{Eq. overdensity}, where $\rho(x)$ is the weighted number of galaxies in the given aperture. 
        To estimate the uncertainty of the weighted galaxy number counts within each aperture, we performed a Monte Carlo sampling by placing 1000 random apertures across the field and computing the weighted counts in each. 
        The distribution of counts from these random apertures reflects the expected background fluctuations due to shot noise. 
        The 16th and 84th percentiles of this non-Gaussian distribution are adopted as the lower and upper bounds of the 68\% confidence interval. 
        Suppose the measured count within the target aperture lies outside this interval. In that case, we adopt a one-sided error bar by setting the error in the direction beyond the sampled distribution to zero, thereby avoiding unphysical or misleading uncertainties.
        \begin{equation}
            \label{Eq. overdensity}
            \delta = \frac{\rho(x)-\bar{\rho}}{\bar{\rho}} = \frac{\rho(x)}{\bar{\rho}} - 1
        \end{equation}
        Where $\Bar{\rho}$ represents the average number density measured from 1000 random positions across the field at the same redshift as the central galaxy.
    
        \subsubsection{Significance of Clustering}
        \label{Sigma & Radius}
            For the previous search of protoclusters, many spectroscopically-selected galaxies are targeted as candidates, which have a small physical separation (< 3 cMpc). However, according to the Millennium-based simulations, protoclusters would have different physical sizes \citep{2022Yajima, Chiang2017}, which are related to the final mass at $z = 0$ \citep{2015Muldrew, 2018Lovell}. Such physical distances only include the very centre, or core, of the protocluster as such a distance universe.
            
            This leads to an analysis of the effect of radius on investigating a range of possible scales of protoclusters and introduces significance ($\sigma$, Eq. \ref{Eq. sigma}) as a quantity to estimate the robustness of each overdensity region that would need to stand out above field fluctuations.
            
            Therefore, we explore different sizes of apertures (R = 0.5 - 7.5 cMpc) for a complete characterisation of protocluster environments at such high redshift. We adopt Eq.15 from \citet{Li2025} for the definition of significance for a certain overdensity value.
            It is defined as the ratio between the difference of overdensity and its average ($\mu_{\delta}$) to the standard deviation ($\sigma_{\delta}$) of the distribution of the overdensity measured from the identical Monte Carlo sampling mentioned in Sec.\ref{weighted Num.-Den.} at the same redshift as the central galaxy.
            
            \begin{equation}
                \label{Eq. sigma}
                \sigma = \frac{\delta - \mu_{\delta}}{\sigma_{\delta}}
            \end{equation}
            Because each galaxy contributes a continuous membership probability, the count distribution follows a Poisson-binomial law. We therefore estimate uncertainties by the distribution from those random apertures and quote the difference between the 84th percentile and median as symmetric errors.
\section{Results and Discussion}
\label{Results}
    \subsection{The Impact of Aperture Size on Overdensity Measurements}
\begin{figure*}
            \centering
            \includegraphics[width=0.48\linewidth]{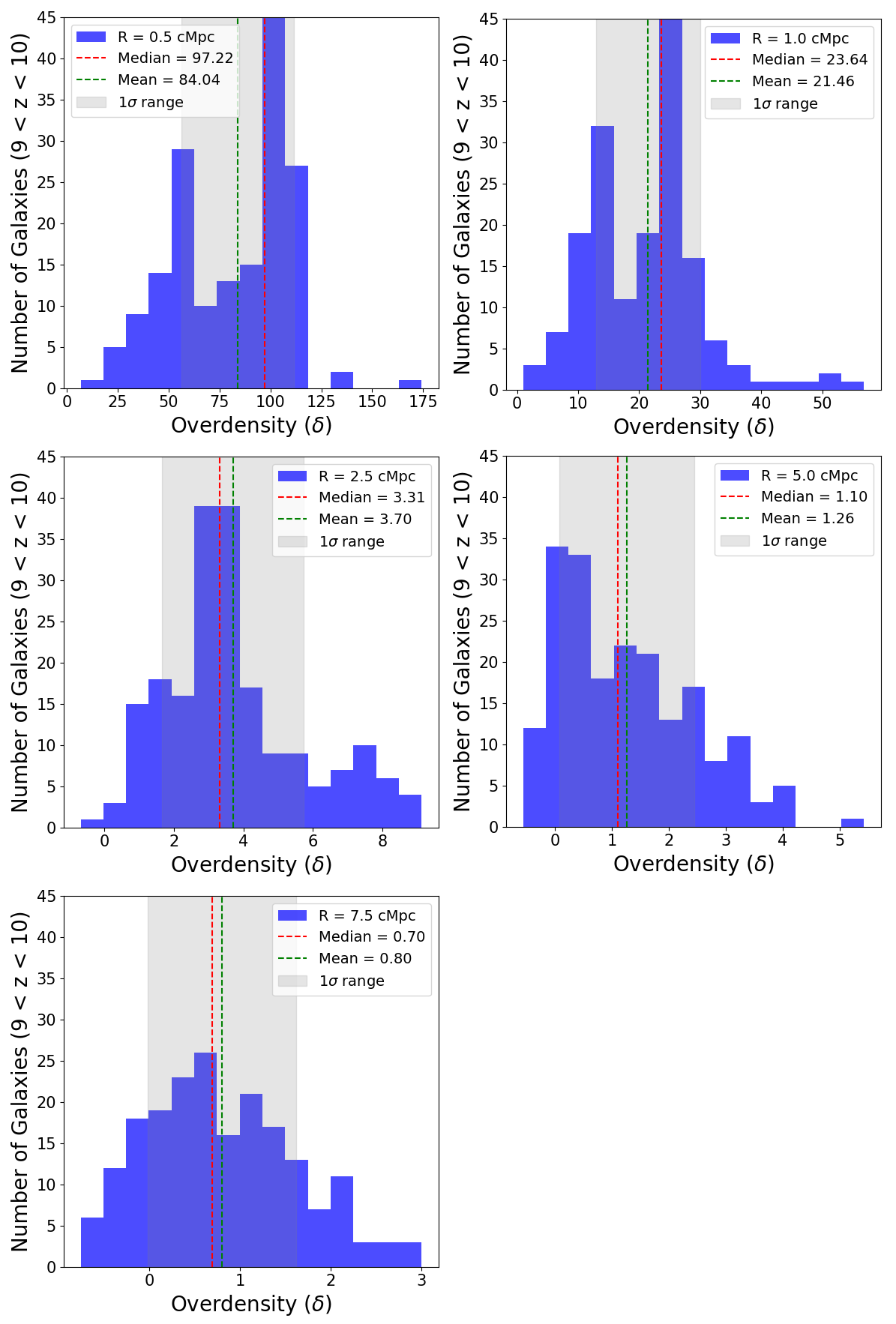}
            \includegraphics[width=0.48\linewidth]{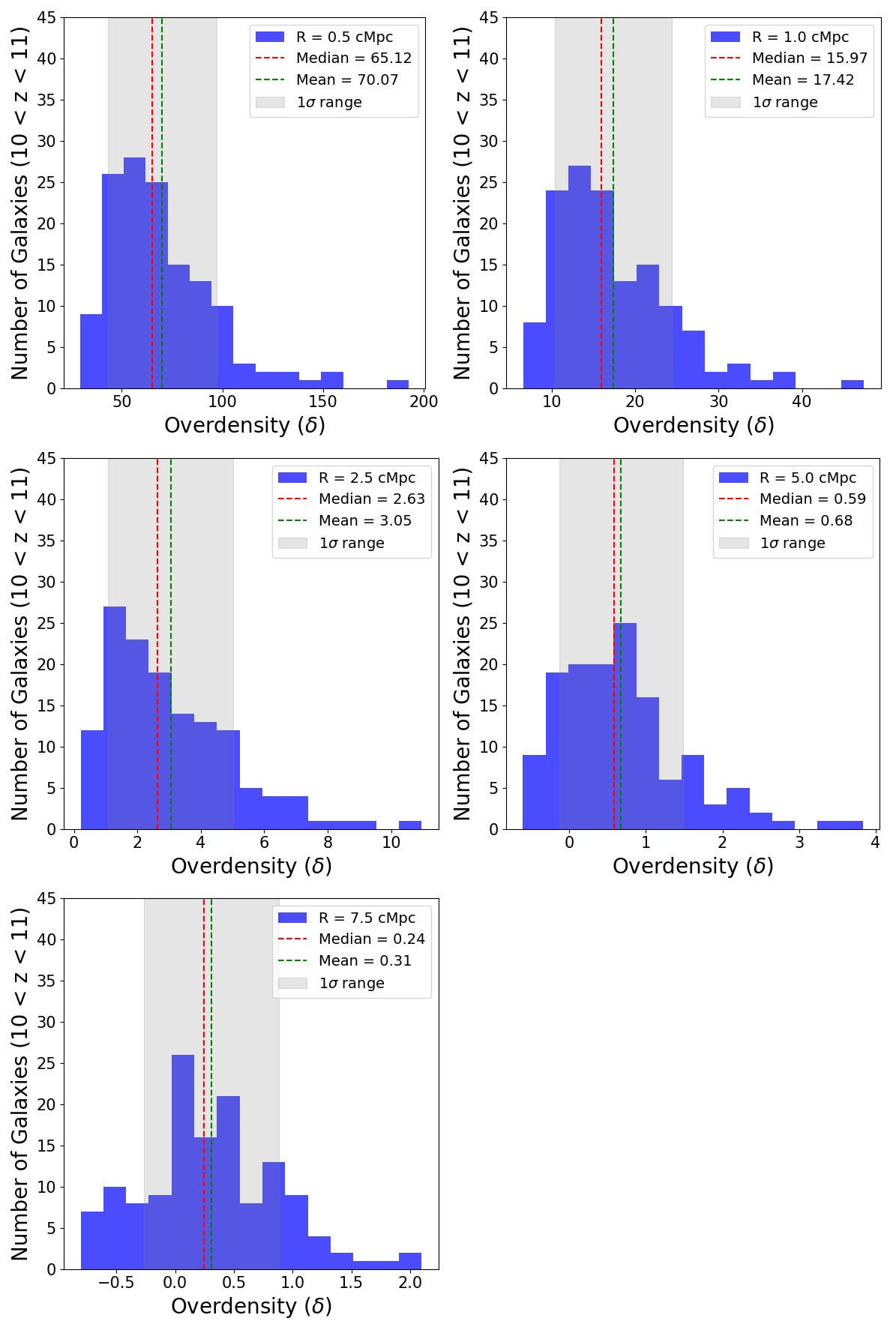}
            \caption{Histograms of galaxy overdensity, $\delta$, measured for all galaxies in the field at different aperture radii, R = 0.5, 1.0, 2.5, 5.0, and 7.5 cMpc (from top to bottom, left to right), for two redshift intervals: (Left) $9 < z < 10$, (Right) $10 < z < 11$.
            Each panel shows the distribution of $\delta$ values, with the blue bars representing the number of galaxies. The shaded region indicates the 1$\sigma$ range around the mean. The red dashed line marks the median, while the green dashed line shows the mean overdensity. The legend in each panel indicates the aperture size and the corresponding statistical values.
            Note that the overdensity distribution becomes narrower and shifts toward lower values as the aperture radius increases, reflecting the dilution of local enhancements over larger spatial scales.}
            \label{fig:overdensity_hist}
        \end{figure*}

        To quantitatively assess how aperture size influences overdensity measurements, we calculated $\delta$ for each galaxy using five representative radii: R =0.5, 1.0, 2.5, 5.0, and 7.5 cMpc. These choices span the observed range of protocluster core sizes, from the very compact A2744z8OD \citep[$\sim$0.3 cMpc,$\delta\approx130$][]{Ishigaki2016} to more extended structures with physical sizes of $2-8$ cMpc at $z\sim8$ \citep{Trenti2012, Helton2024b, Li2025}, as well as the typical half-mass radii in simulations \citep[$\lesssim 10$ cMpc][]{Chiang2013}. The resulting overdensity distributions for each aperture size are shown in Figure \ref{fig:overdensity_hist}.
        
        As seen in Figure \ref{fig:overdensity_hist}, the overdensity distributions for smaller apertures (R = 0.5 and 1.0 cMpc) are strongly skewed toward high $\delta$ values, with medians of $\delta \approx$ 97.22 and 23.64, respectively. This is primarily due to small sampling volumes, which result in extreme $\delta$ values ($\delta \gtrsim$ 20). In contrast, when using larger apertures (R = 2.5, 5.0, and 7.5 cMpc), the median $\delta$ values approach zero ($\delta \approx$ 3.31, 1.10, and 0.70). With the even larger radius, 7.5 cMpc, all the densities are diluted to below 4 times denser than the field average ($\delta < 3$).

        Among these, we therefore found that a core aperture radius of 2.5 cMpc offers a balanced sensitivity: it is large enough to encompass significant physical structure, while still focusing on the densest core regions where the protocluster signature is most robust. Adopting smaller apertures tends to fragment physically connected structures, while larger apertures dilute the overdensity signal due to increasing inclusion of the surrounding field. Therefore, 2.5 cMpc is chosen as an optimal compromise for the identification of protocluster cores in our dataset, and we have adopted R = 2.5 cMpc as the fiducial radius for calculating the local-/overdensity value on each galaxy.

    \subsection{Protocluster Candidates}
        First, we identify the 'core' galaxies of protocluster candidates as those with overdensity ($\delta$) values exceeding the 95th percentile of the $\delta$ distribution for all galaxies and with statistically significant peaks based on Monte Carlo sampling ($\sigma > 3$). To assemble full protocluster candidates, we link galaxies located within a projected distance of 7.5 cMpc from these cores. \citet{Chiang2017} demonstrated that the probability of a galaxy being associated with the protocluster core drops sharply beyond this scale at $z \sim 7$, which is why we use this linking length.

        Additionally, if two core galaxies are within 7.5 cMpc, we merge them into a single larger structure. Due to the larger radius used to search for members (7.5 cMpc), we can still detect large structures, as demonstrated with R = 5.0 or 7.5 cMpc. The reported core candidates and members are identified at different scales to maximise our sensitivity to a range of protocluster morphologies and to ensure comparability with previous studies using various methodologies.
        
\begin{figure*}
        \centering
        \includegraphics[width=0.8\columnwidth]{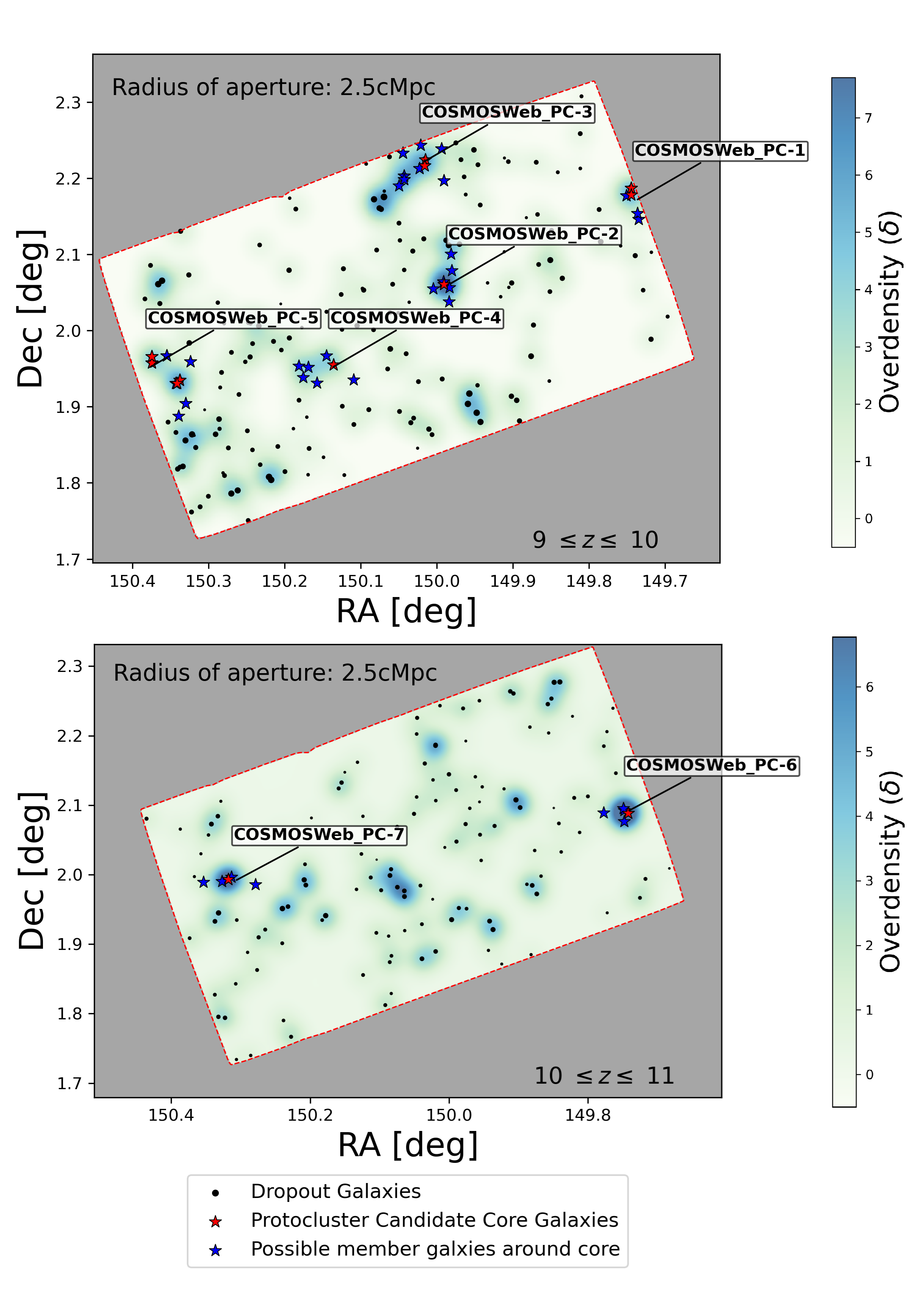}
        \caption{This shows the projected distribution of F115W dropout galaxies across the COSMOS-Web footprint. Black circles mark all sources whose best-fit photometric redshifts satisfy the range $9\le z_{\mathrm{best}}\le10$ (or $10\le z_{\mathrm{best}}\le11$ in the bottom panel). The size of each scatter point represents the significance ($\sigma$) of its overdensity value compared to the Monte Carlo sampling. A Gaussian-kernel surface-density map (shown in green-to-blue shading) highlights significant overdensities. Within each overdense peak, galaxies with overdensities greater than the 95th percentile of the entire distribution ($\delta\ge\delta_{95}$) and robust enough ($\sigma>3$) over the Monte Carlo sampling are designated a protocluster core candidate (red star), while blue stars indicate galaxies lying inside an R=7.5 cMpc aperture centred on the cores. The red polygon outlines the NIRCam field of view, and the grey background denotes regions outside the imaging coverage.}
        \label{fig:2DScatter}
    \end{figure*}

        Figure \ref{fig:2DScatter} shows the most significant photometrically selected protocluster candidates in the COSMOSWeb DR 0.5 field of view. For illustrative purposes, we convolved a Gaussian kernel density estimation (KDE) with the overdensity colourmap. The $1\sigma$ width of the Gaussian kernel is set to 2.5 cMpc. The top panel shows the five most promising photometrically selected protocluster candidates in $9 \leq z \leq 10$, labelled \texttt{COSMOSWeb\_PC-1}, \texttt{COSMOSWeb\_PC-2}, \texttt{COSMOSWeb\_PC-3}, \newline \texttt{COSMOSWeb\_PC-4}, and \texttt{COSMOSWeb\_PC-5}. The bottom panel shows the two significant detections of protocluster candidates, \texttt{COSMOSWeb\_PC-6} and \texttt{COSMOSWeb\_PC-7}, at $10 \leq z \leq 11$.

        The detailed false-colour cutouts and the probability density function (PDF) for the members of each protocluster are shown in the left panel of Figs. \ref{fig:PC1}-\ref{fig:PC7}. Solid PDFs are more likely to be associated with the central galaxy (p > 50\%), whereas the galaxies have a lower probability (2.5\% < p < 50\%) of being associated with the core galaxies indicated by dashed lines (corresponding to $<2\sigma$ if assuming a Gaussian distribution). The spatial clustering and photometric redshift coherence suggest that these five highlighted groups are promising protoclusters at cosmic dawn.
        We summarise the position, overdensity, significance, photometry, and stellar mass of each member of the reported protocluster candidates in Tables \ref{tab:PC_summary} and \ref{tab:PC_summary2}.
        
        \subsubsection{Physical Size of Protocluster Candidates}
            The physical size of the forming galaxy (protocluster), i.e. the region expected to contain at least 50\% of the descendant mass according to the half-mass radius statistics of cosmological simulations \citep{Chiang2013}, is suggested to be less than 10 cMpc. However, recent searches for protoclusters are mainly limited by the survey area, which could lead to the larger half-mass radius size of 7-10 cMpc being incomplete. The overdensity reported by \citet{Helton2024b} was obtained within a volume of 15 cMpc$^3$, with an NIRCam Wide Field Slitless Spectroscopy (WFSS) survey that covered only $\approx 62$ arcmin$^2$. Two overdensities containing four and six galaxies were found at $z \sim 8$, with physical sizes of 3.9 and 5.8 cMpc, respectively. \citet{Trenti2012} also identified an overdensity ($\sigma \approx$ 8.9, relative to the field mean) at $z \approx 8$ in the BoRG survey, using a search box of $2.83 \times 2.83$ cMpc$^2$ (corresponding to 62" $\times$ 62" at $z = 8.0$). By contrast, our search method benefits from a wider survey area. Using an aperture radius of 7.5 cMpc, we can cover larger structures with sizes of $\sim15$ cMpc.
            
\begin{figure*}
        \centering
        \includegraphics[width=\columnwidth]{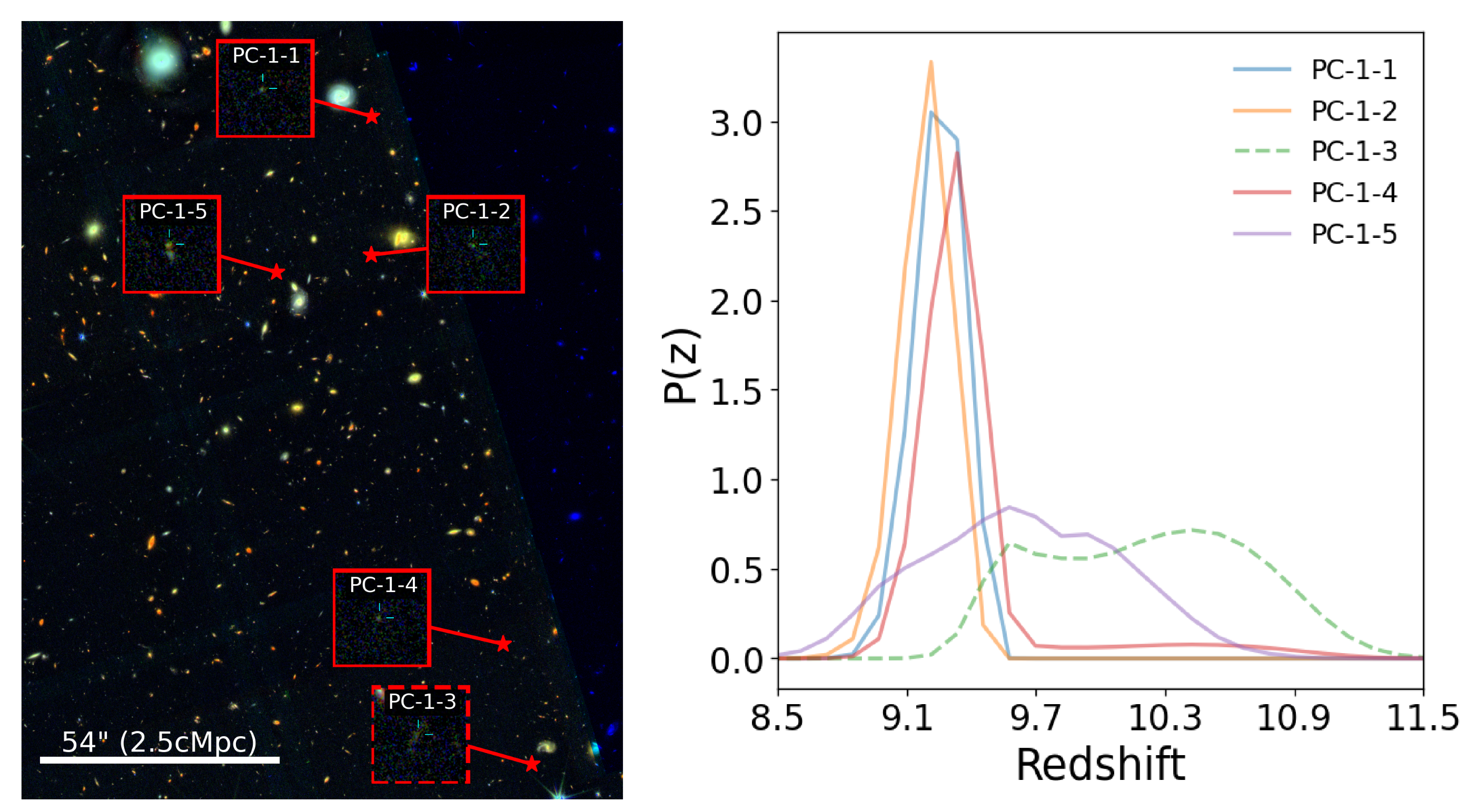}
        \caption{\textbf{Left:} \textit{JWST}/NIRCam three–colour mosaic (B: F814W+F115W, G:F150W+F277W, R:F444W) centred on the COSMOS-Web protocluster candidate COSMOS-Web\_PC-1 ($z_{\mathrm{phot}}\simeq 9.3$).  
        Red stars mark the positions of member/core galaxies. Solid frames identify candidate members that have higher probabilities ($W>0.5$) associated with the cores, while dashed frames mark lower probabilities ($0.025< W \leq0.5$) sources. Each cutout has a Field-of-View of 4"$\times$4". The white bar in the lower-left corner corresponds to 54" (2.5 cMpc) at $z\simeq 9.5$. North is up, and east is to the left.
        \textbf{Right:} The PDFs of cores/potential members for the protocluster candidate COSMOS-Web\_PC-1. Solid/Dashed lines denote those solid/dashed frame sources in the left panel.}
        \label{fig:PC1}
    \end{figure*}

\begin{figure*}
        \centering
        \includegraphics[width=\columnwidth]{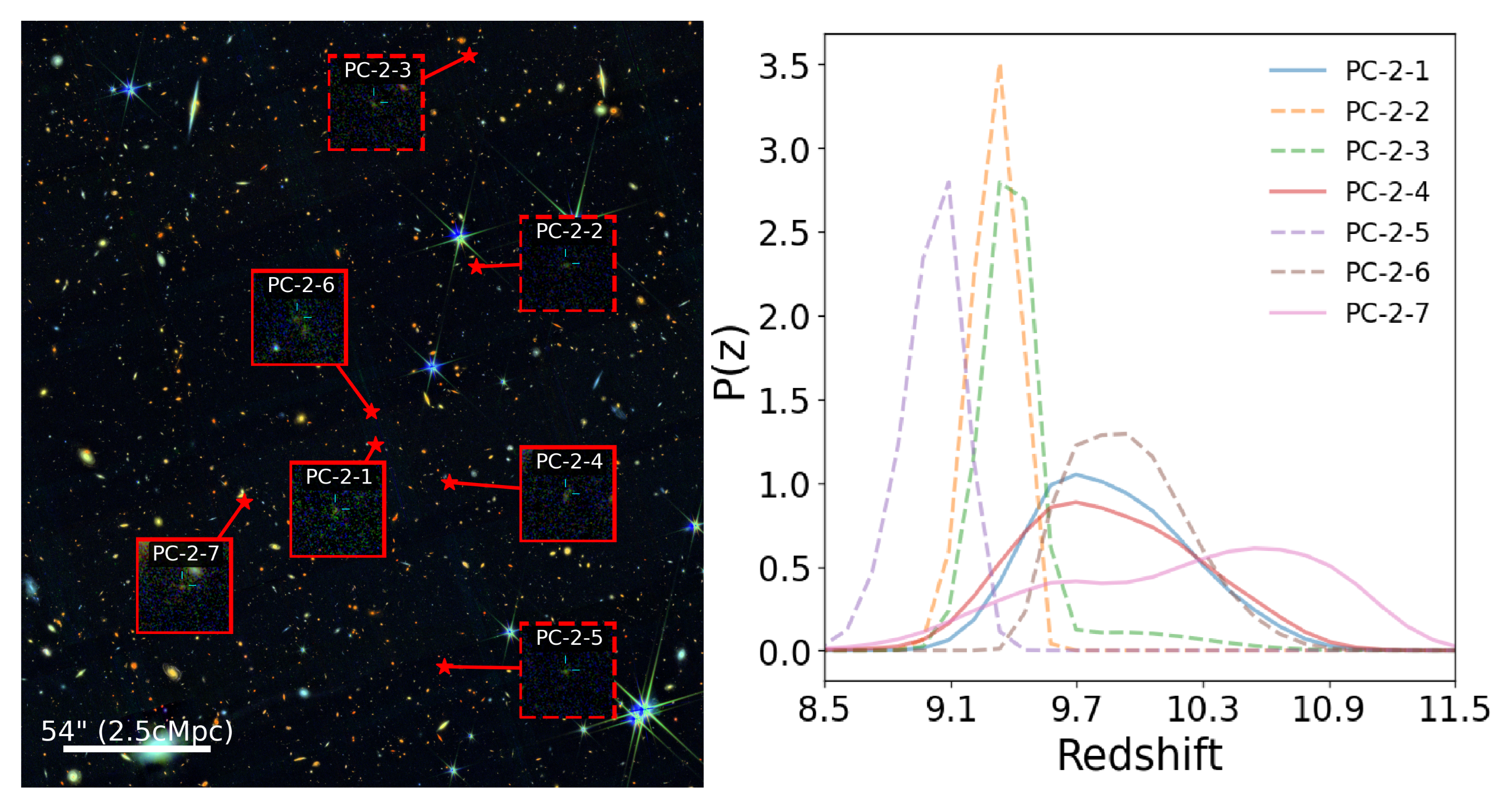}
        \caption{Same as Figure\ref{fig:PC1}, but for COSMOSWeb\_PC-2.}
        \label{fig:PC2}
    \end{figure*}

\begin{figure*}
        \centering
        \includegraphics[width=\columnwidth]{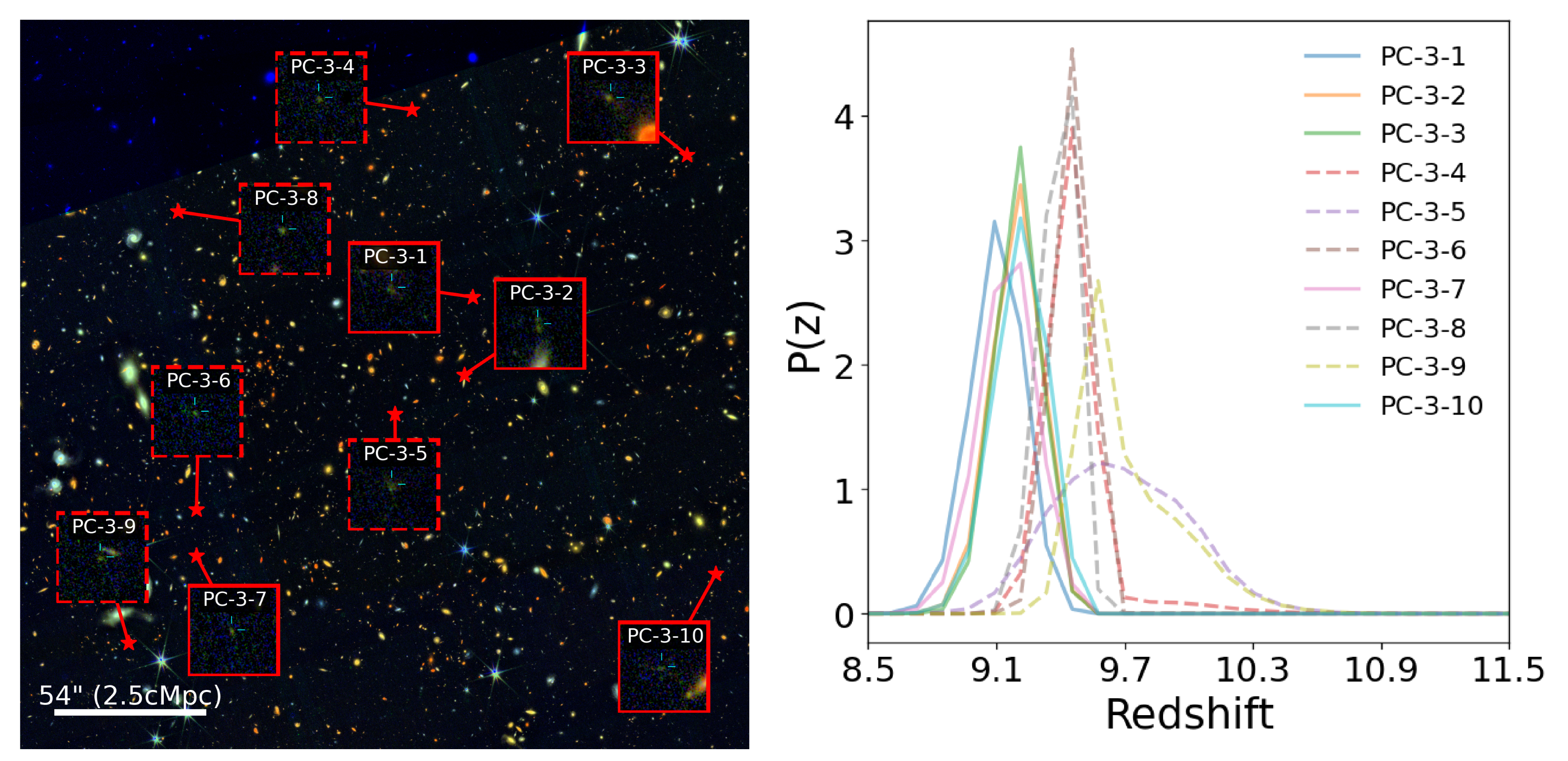}
        \caption{Same as Figure\ref{fig:PC1}, but for COSMOSWeb\_PC-3.}
        \label{fig:PC3}
    \end{figure*}

\begin{figure*}
        \centering
        \includegraphics[width=\columnwidth]{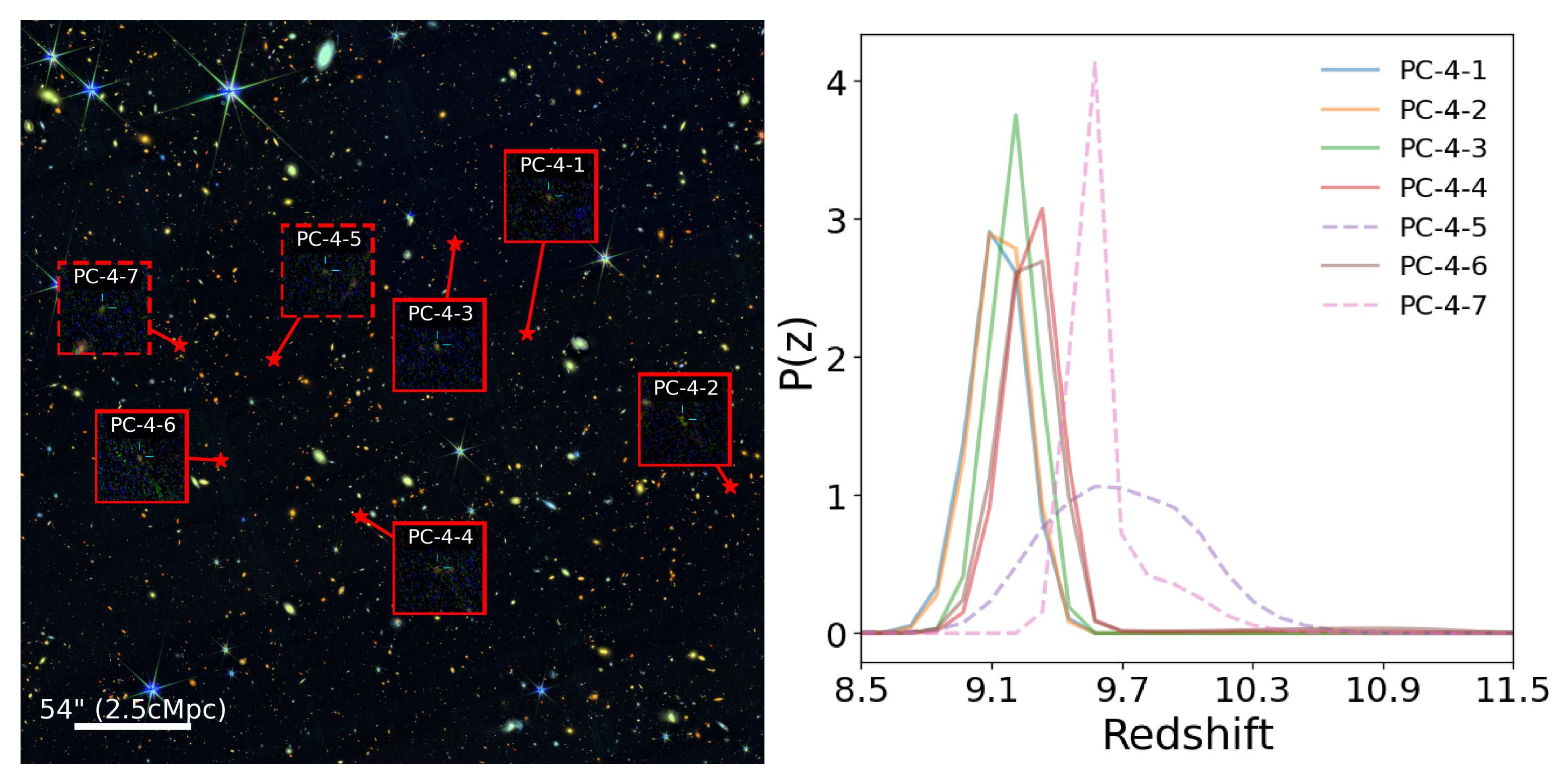}
        \caption{Same as Figure\ref{fig:PC1}, but for COSMOSWeb\_PC-4.}
        \label{fig:PC4}
    \end{figure*}

\begin{figure*}
        \centering
        \includegraphics[width=\columnwidth]{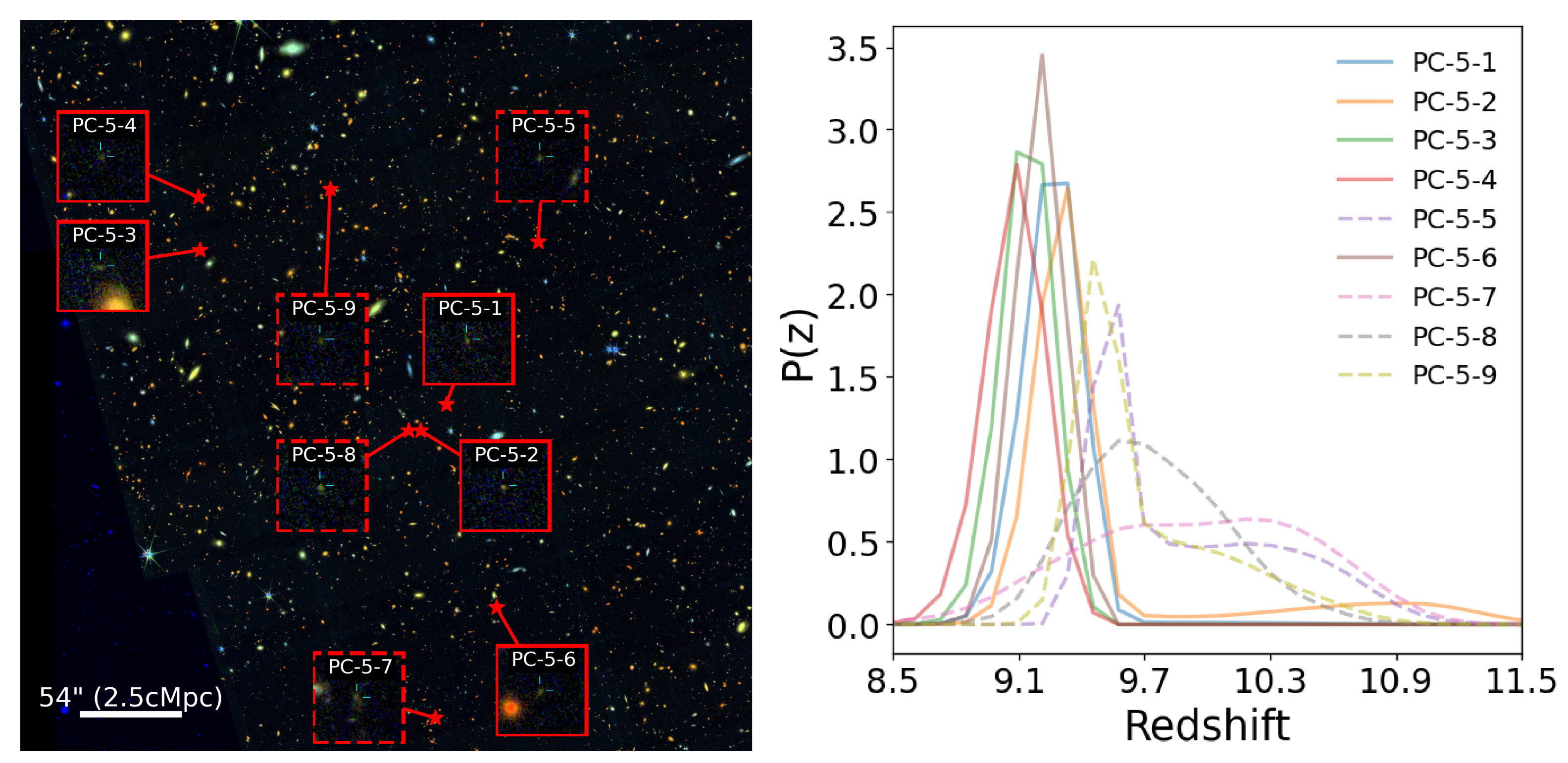}
        \caption{Same as Figure\ref{fig:PC1}, but for COSMOSWeb\_PC-5.}
        \label{fig:PC5}
    \end{figure*}

\begin{figure*}
        \centering
        \includegraphics[width=\columnwidth]{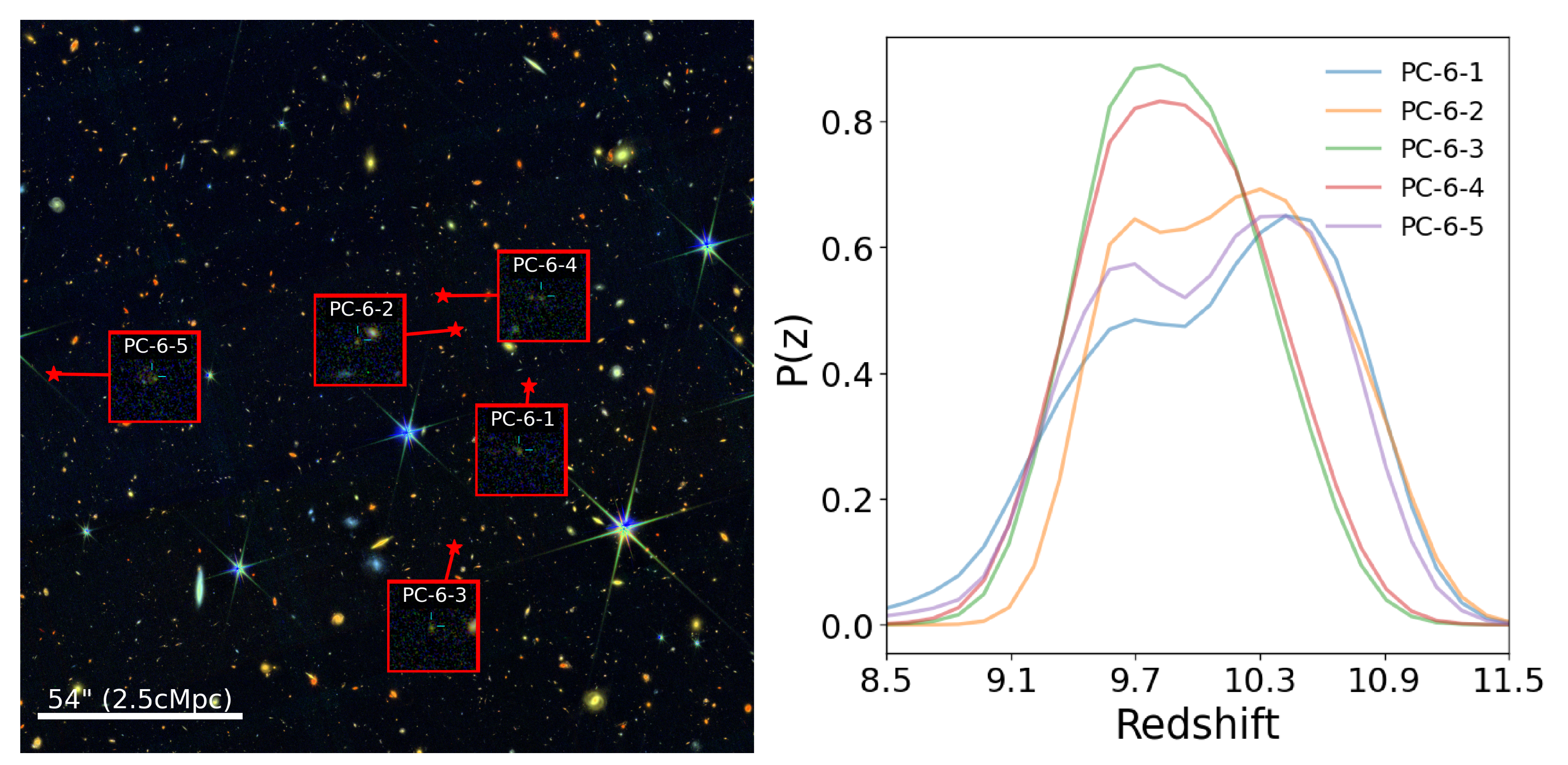}
        \caption{Same as Figure\ref{fig:PC1}, but for COSMOSWeb\_PC-6.}
        \label{fig:PC6}
    \end{figure*}

\begin{figure*}
        \centering
        \includegraphics[width=\columnwidth]{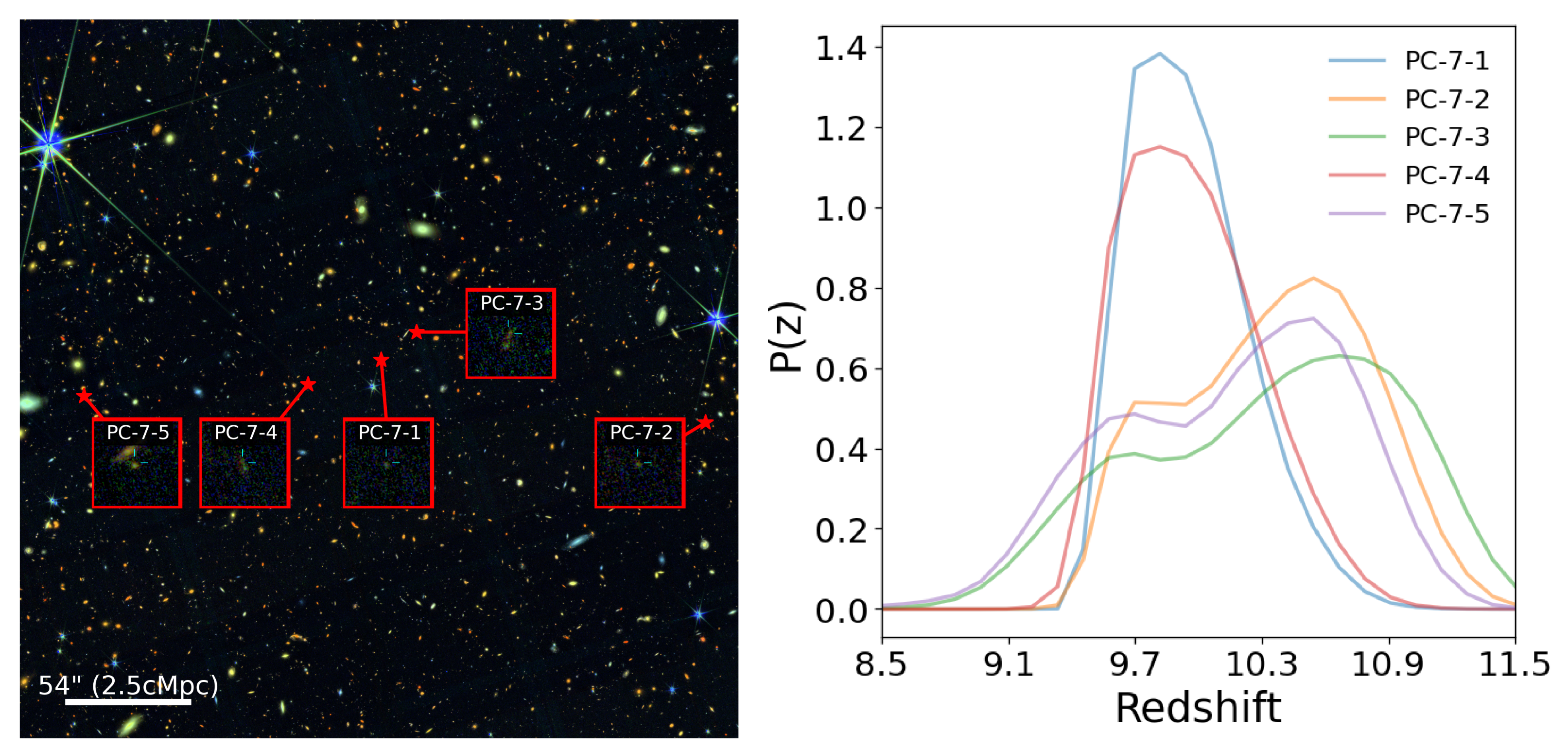}
        \caption{Same as Figure\ref{fig:PC1}, but for COSMOSWeb\_PC-7.}
        \label{fig:PC7}
    \end{figure*}

            We estimate the size of each protocluster candidate by identifying the members that are most separated in right ascension and declination, with depth defined as the maximum difference in best-fit redshifts among highly associated members.
            \texttt{COSMOSWeb\_PC-1} ($\delta_{\rm max} = 9.04$, $\sigma_{\rm max} = 4.27$, Figure \ref{fig:PC1}), this structure comprises five galaxies at $z \sim 9.3$. The overdensity extends across a volume of approximately $2.8 \times 7.4 \times 243.7$ cMpc.
            \texttt{COSMOSWeb\_PC-2} ($\delta_{\rm max} = 9.13$, $\sigma_{\rm max} = 3.65$, Figure \ref{fig:PC2}) contains eight galaxies at $z \sim 9.9$. We estimate the total volume of this protocluster to be $5.5 \times 12.0 \times 226.8$ cMpc. 
            \texttt{COSMOSWeb\_PC-3} ($\delta_{\rm max} = 8.73$, $\sigma_{\rm max} = 4.20$, Figure \ref{fig:PC3}) contains ten galaxies at $z \sim 9.2$. We estimate the total volume of this protocluster to be $11.0 \times 9.1 \times 241.9$ cMpc. 
            \texttt{COSMOSWeb\_PC-4} ($\delta_{\rm max} = 8.12$, $\sigma_{\rm max} = 3.63$, Figure \ref{fig:PC4}) contains seven galaxies at $z \sim 9.2$. We estimate the total volume of this protocluster to be 14.7 x 6.4 x 358.2 cMpc. 
            \texttt{COSMOSWeb\_PC-5} ($\delta_{\rm max} = 8.56$, $\sigma_{\rm max} = 4.41$, Figure \ref{fig:PC5}) contains nine galaxies at $z \sim 9.3$. We estimate the total volume of this protocluster to be $9.3 \times 16.6 \times 225.2$ cMpc. 
            \texttt{COSMOSWeb\_PC-6} ($\delta_{\rm max} = 10.96$, $\sigma_{\rm max} = 3.78$, Figure \ref{fig:PC6}) contains five galaxies at $z \sim 10.3$. We estimate the total volume of this protocluster to be $8.2 \times 4.0 \times 206.2$ cMpc. 
            \texttt{COSMOSWeb\_PC-7} ($\delta_{\rm max} = 8.80$, $\sigma_{\rm max} = 3.16$, Figure \ref{fig:PC7}) contains five galaxies at $z \sim 10.0$. We estimate the total volume of this protocluster to be $13.1 \times 1.9 \times 211.7$ cMpc.

    
    \begin{table*}
        \centering
        \begin{tabular}{ccccccccccc}
            ID & RA & Dec & $\delta$ & $\sigma$ & Best-fit $\chi^{2}$ & $z_{\text{best}}$ & $F_{\text{F115W}}$ [nJy] & $F_{\text{F150W}}$ [nJy] & Stellar Mass (log($M/M_{\odot}$)) \\ [1ex] \hline
            PC-1-1 & 09h58m58.77s & +02d11m14.20s & $8.26^{+5.22}_{-5.22}$ & 4.26 & 0.09 & 9.29 & $8.98\pm7.93$ & $44.35\pm4.62$ & 8.04 $^{+0.04}_{-0.05}$ \\ [1ex]
            PC-1-2 & 09h58m58.78s & +02d10m42.64s & $9.04^{+5.66}_{-5.66}$ & 4.27 & 0.54 & 9.17 & $10.99\pm8.75$ & $45.55\pm4.75$ & 8.06 $^{+0.04}_{-0.05}$ \\ [1ex]
            PC-1-3 & 09h58m56.34s & +02d08m46.83s & $2.15^{+0.44}_{-0.44}$ & 1.20 & 0.09 & 9.53 & $1.09\pm7.74$ & $32.14\pm3.57$ & 8.26 $^{+0.04}_{-0.05}$ \\ [1ex]
            PC-1-4 & 09h58m56.78s & +02d09m14.12s & $3.03^{+0.32}_{-0.32}$ & 1.61 & 0.09 & 9.35 & $4.53\pm7.89$ & $32.08\pm3.55$ & 7.96 $^{+0.04}_{-0.05}$ \\ [1ex]
            PC-1-5 & 09h59m00.22s & +02d10m38.78s & $2.07^{+0.65}_{-0.65}$ & 1.15 & 0.78 & 9.59 & $11.26\pm56.30$ & $84.38\pm8.59$ & 9.17 $^{+0.04}_{-0.05}$ \\ [1ex]
            \hline
            PC-2-1 & 09h59m57.83s & +02d03m38.75s & $9.13^{+7.63}_{-7.63}$ & 3.65 & 0.17 & 9.95 & $5.99\pm29.97$ & $51.91\pm5.44$ & 7.72 $^{+0.04}_{-0.05}$ \\ [1ex]
            PC-2-2 & 09h59m55.33s & +02d04m44.83s & $3.49^{+0.79}_{-0.79}$ & 1.86 & 0.14 & 9.35 & $5.12\pm5.41$ & $32.59\pm3.40$ & 8.23 $^{+0.04}_{-0.05}$ \\ [1ex]
            PC-2-3 & 09h59m55.51s & +02d06m03.18s & $4.61^{+2.24}_{-2.24}$ & 2.52 & 0.09 & 9.41 & $3.95\pm7.17$ & $33.90\pm3.66$ & 8.45 $^{+0.04}_{-0.05}$ \\ [1ex]
            PC-2-4 & 09h59m56.00s & +02d03m24.72s & $5.92^{+4.67}_{-4.67}$ & 3.08 & 0.18 & 9.71 & $7.05\pm35.26$ & $38.17\pm4.13$ & 7.62 $^{+0.04}_{-0.05}$ \\ [1ex]
            PC-2-5 & 09h59m56.12s & +02d02m16.28s & $3.74^{+0.50}_{-0.50}$ & 1.56 & 0.80 & 9.05 & $13.86\pm8.18$ & $47.16\pm4.80$ & 8.05 $^{+0.04}_{-0.05}$ \\ [1ex]
            PC-2-6 & 09h59m57.93s & +02d03m51.04s & $7.02^{+5.51}_{-5.51}$ & 2.81 & 0.17 & 9.95 & $1.31\pm6.56$ & $41.57\pm4.30$ & 8.37 $^{+0.04}_{-0.05}$ \\ [1ex]
            PC-2-7 & 10h00m01.08s & +02d03m17.49s & $2.36^{+0.99}_{-0.99}$ & 1.09 & 0.17 & 9.83 & $6.86\pm34.32$ & $36.76\pm3.98$ & 8.42 $^{+0.04}_{-0.05}$ \\ [1ex]
            \hline
            PC-3-1 & 10h00m03.68s & +02d13m28.91s & $8.73^{+5.35}_{-5.35}$ & 4.13 & 0.09 & 9.17 & $12.73\pm7.87$ & $46.22\pm4.71$ & 7.75 $^{+0.04}_{-0.05}$ \\ [1ex]
            PC-3-2 & 10h00m03.88s & +02d13m00.89s & $8.16^{+4.90}_{-4.90}$ & 4.04 & 0.12 & 9.23 & $10.18\pm7.01$ & $44.20\pm4.52$ & 8.09 $^{+0.04}_{-0.05}$ \\ [1ex]
            PC-3-3 & 09h59m58.53s & +02d14m20.05s & $3.50^{+0.25}_{-0.25}$ & 1.74 & 0.10 & 9.23 & $8.22\pm6.40$ & $42.73\pm4.38$ & 8.11 $^{+0.04}_{-0.05}$ \\ [1ex]
            PC-3-4 & 10h00m05.14s & +02d14m36.37s & $3.36^{+1.38}_{-1.38}$ & 1.87 & 0.09 & 9.47 & $4.19\pm7.12$ & $44.53\pm4.66$ & 8.11 $^{+0.04}_{-0.05}$ \\ [1ex]
            PC-3-5 & 10h00m05.55s & +02d12m46.70s & $2.03^{+0.70}_{-0.70}$ & 1.10 & 0.20 & 9.65 & $14.13\pm70.65$ & $52.88\pm5.37$ & 8.18 $^{+0.04}_{-0.05}$ \\ [1ex]
            PC-3-6 & 10h00m10.32s & +02d12m12.39s & $7.55^{+5.57}_{-5.57}$ & 4.20 & 0.89 & 9.47 & $4.00\pm4.99$ & $51.25\pm5.21$ & 8.10 $^{+0.04}_{-0.05}$ \\ [1ex]
            PC-3-7 & 10h00m10.32s & +02d11m55.66s & $5.68^{+2.30}_{-2.30}$ & 2.68 & 0.77 & 9.17 & $8.96\pm8.57$ & $46.30\pm4.85$ & 8.69 $^{+0.04}_{-0.05}$ \\ [1ex]
            PC-3-8 & 10h00m10.77s & +02d13m59.68s & $4.59^{+2.22}_{-2.22}$ & 2.51 & 0.14 & 9.41 & $7.60\pm7.34$ & $63.66\pm6.49$ & 7.79 $^{+0.04}_{-0.05}$ \\ [1ex]
            PC-3-9 & 10h00m11.95s & +02d11m24.32s & $7.11^{+5.40}_{-5.40}$ & 3.98 & 0.38 & 9.53 & $2.16\pm6.83$ & $47.74\pm4.96$ & 7.95 $^{+0.04}_{-0.05}$ \\ [1ex]
            PC-3-10 & 09h59m57.83s & +02d11m49.16s & $2.89^{+0.14}_{-0.14}$ & 1.49 & 0.22 & 9.29 & $8.98\pm7.91$ & $42.82\pm4.48$ & 8.00 $^{+0.04}_{-0.05}$ \\ [1ex]
            \hline
            PC-4-1 & 10h00m32.60s & +01d57m19.11s & $8.12^{+4.82}_{-4.82}$ & 3.63 & 0.09 & 9.11 & $9.66\pm8.33$ & $40.48\pm4.27$ & 8.37 $^{+0.04}_{-0.05}$ \\ [1ex]
            PC-4-2 & 10h00m26.22s & +01d56m06.97s & $2.94^{+0.31}_{-0.31}$ & 1.46 & 0.11 & 9.23 & $10.92\pm7.62$ & $43.19\pm4.45$ & 7.90 $^{+0.04}_{-0.05}$ \\ [1ex]
            PC-4-3 & 10h00m34.86s & +01d58m01.23s & $7.18^{+3.92}_{-3.92}$ & 3.55 & 0.10 & 9.23 & $10.21\pm6.84$ & $51.54\pm5.23$ & 8.12 $^{+0.04}_{-0.05}$ \\ [1ex]
            PC-4-4 & 10h00m37.82s & +01d55m53.02s & $3.00^{+0.04}_{-0.04}$ & 1.55 & 0.09 & 9.29 & $5.07\pm7.42$ & $34.81\pm3.75$ & 8.34 $^{+0.04}_{-0.05}$ \\ [1ex]
            PC-4-5 & 10h00m40.53s & +01d57m06.74s & $5.99^{+4.66}_{-4.66}$ & 2.95 & 0.17 & 9.77 & $11.03\pm55.16$ & $38.30\pm3.94$ & 8.07 $^{+0.04}_{-0.05}$ \\ [1ex]
            PC-4-6 & 10h00m42.20s & +01d56m19.23s & $5.13^{+1.88}_{-1.88}$ & 2.54 & 0.10 & 9.23 & $4.87\pm7.33$ & $31.81\pm3.47$ & 8.39 $^{+0.04}_{-0.05}$ \\ [1ex]
            PC-4-7 & 10h00m43.49s & +01d57m13.55s & $5.27^{+3.56}_{-3.56}$ & 2.95 & 0.15 & 9.53 & $2.32\pm4.83$ & $49.36\pm5.02$ & 8.14 $^{+0.04}_{-0.05}$ \\ [1ex]
            \hline
            PC-5-1 & 10h01m21.03s & +01d56m04.86s & $8.56^{+5.52}_{-5.52}$ & 4.41 & 0.09 & 9.29 & $6.34\pm9.38$ & $40.11\pm4.38$ & 8.68 $^{+0.04}_{-0.05}$ \\ [1ex]
            PC-5-2 & 10h01m21.95s & +01d55m50.77s & $8.25^{+5.21}_{-5.21}$ & 4.25 & 0.09 & 9.29 & $5.27\pm8.94$ & $38.64\pm4.22$ & 8.48 $^{+0.04}_{-0.05}$ \\ [1ex]
            PC-5-3 & 10h01m29.88s & +01d57m28.17s & $8.01^{+4.71}_{-4.71}$ & 3.58 & 0.11 & 9.11 & $11.77\pm8.92$ & $46.38\pm4.83$ & 7.94 $^{+0.04}_{-0.05}$ \\ [1ex]
            PC-5-4 & 10h01m29.93s & +01d57m56.66s & $8.01^{+4.71}_{-4.71}$ & 3.58 & 0.09 & 9.11 & $8.46\pm8.07$ & $31.89\pm3.47$ & 8.00 $^{+0.04}_{-0.05}$ \\ [1ex]
            PC-5-5 & 10h01m17.72s & +01d57m32.63s & $1.20^{+0.52}_{-0.52}$ & 0.67 & 0.13 & 9.53 & $1.85\pm4.48$ & $36.51\pm3.75$ & 7.68 $^{+0.04}_{-0.05}$ \\ [1ex]
            PC-5-6 & 10h01m19.21s & +01d54m15.25s & $3.27^{+0.01}_{-0.01}$ & 1.62 & 0.09 & 9.23 & $8.78\pm6.18$ & $40.60\pm4.15$ & 8.28 $^{+0.04}_{-0.05}$ \\ [1ex]
            PC-5-7 & 10h01m21.41s & +01d53m15.76s & $1.10^{+0.29}_{-0.29}$ & 0.47 & 0.17 & 9.89 & $7.10\pm35.50$ & $36.67\pm3.76$ & 8.48 $^{+0.04}_{-0.05}$ \\ [1ex]
            PC-5-8 & 10h01m22.37s & +01d55m50.79s & $4.42^{+3.08}_{-3.08}$ & 2.39 & 0.33 & 9.65 & $15.45\pm77.23$ & $64.09\pm6.57$ & 7.79 $^{+0.04}_{-0.05}$ \\ [1ex]
            PC-5-9 & 10h01m25.20s & +01d58m01.06s & $1.58^{+0.39}_{-0.39}$ & 0.88 & 0.10 & 9.47 & $2.72\pm6.84$ & $35.35\pm3.78$ & 7.58 $^{+0.04}_{-0.05}$ \\ [1ex]
            \hline
        \end{tabular} \\
        \caption{The basic information of the member galaxies in each protocluster candidate.}
        \label{tab:PC_summary}
    \end{table*}
    \begin{table*}
            \centering
            \begin{tabular}{ccccccccccc}
            ID & RA & Dec & $\delta$ & $\sigma$ & Best-fit $\chi^{2}$ & $z_{\text{best}}$ & $F_{\text{F115W}}$ [nJy] & $F_{\text{F150W}}$ [nJy] & Stellar Mass (log($M/M_{\odot}$)) \\ [1ex] \hline
            PC-6-1 & 09h58m58.21s & +02d05m18.57s & $10.96^{+9.59}_{-9.59}$ & 3.78 & 0.17 & 10.13 & $10.79\pm53.93$ & $42.98\pm4.48$ & 8.23 $^{+0.04}_{-0.05}$ \\ [1ex]
            PC-6-2 & 09h58m59.51s & +02d05m33.42s & $6.34^{+4.86}_{-4.86}$ & 2.39 & 0.17 & 10.01 & $3.39\pm16.93$ & $41.86\pm4.37$ & 8.76 $^{+0.04}_{-0.05}$ \\ [1ex]
            PC-6-3 & 09h58m59.53s & +02d04m35.50s & $4.77^{+3.36}_{-3.36}$ & 1.34 & 0.17 & 10.43 & $7.09\pm35.46$ & $51.79\pm5.33$ & 8.01 $^{+0.04}_{-0.05}$ \\ [1ex]
            PC-6-4 & 09h58m59.74s & +02d05m42.49s & $7.67^{+6.38}_{-6.38}$ & 2.36 & 0.17 & 10.31 & $4.97\pm24.84$ & $34.77\pm3.72$ & 8.29 $^{+0.04}_{-0.05}$ \\ [1ex]
            PC-6-5 & 09h59m06.64s & +02d05m21.62s & $1.55^{+0.00}_{-0.00}$ & 0.41 & 0.57 & 10.49 & $11.01\pm55.05$ & $41.13\pm4.33$ & 7.73 $^{+0.04}_{-0.05}$ \\ [1ex]
            \hline
            PC-7-1 & 10h01m16.26s & +01d59m35.84s & $8.80^{+7.26}_{-7.26}$ & 3.16 & 0.17 & 10.07 & $0.76\pm3.80$ & $39.55\pm4.19$ & 7.99 $^{+0.04}_{-0.05}$ \\ [1ex]
            PC-7-2 & 10h01m06.85s & +01d59m08.51s & $1.83^{+0.52}_{-0.52}$ & 0.54 & 0.17 & 10.37 & $1.10\pm5.52$ & $34.25\pm3.78$ & 7.88 $^{+0.04}_{-0.05}$ \\ [1ex]
            PC-7-3 & 10h01m15.24s & +01d59m47.87s & $6.46^{+5.08}_{-5.08}$ & 2.23 & 0.17 & 10.13 & $5.57\pm27.85$ & $39.50\pm4.21$ & 8.50 $^{+0.04}_{-0.05}$ \\ [1ex]
            PC-7-4 & 10h01m18.39s & +01d59m25.18s & $6.86^{+5.57}_{-5.57}$ & 2.11 & 0.17 & 10.31 & $2.05\pm10.25$ & $43.62\pm4.58$ & 8.24 $^{+0.04}_{-0.05}$ \\ [1ex]
            PC-7-5 & 10h01m24.89s & +01d59m20.14s & $1.99^{+0.41}_{-0.41}$ & 0.50 & 0.24 & 10.55 & $14.95\pm74.76$ & $67.56\pm6.82$ & 7.96 $^{+0.04}_{-0.05}$ \\ [1ex]
            \hline
            \end{tabular} \\
            \caption{Continued.}
            \label{tab:PC_summary2}
        \end{table*}

    \subsection{Stellar/Halo Mass of Candidates}
    \label{S/HM of Cand.}
        It is important to note that the linking length used to associate protocluster members and the aperture used to define protocluster cores do not have to be the same. Therefore, when interpreting the data physically, such as when estimating the mass or rarity of protoclusters, it is important to use a radius comparable to the expected size of the protocluster at the relevant redshift ($z > 9$). Here, we adopt a radius of 7.5 cMpc when summing up the stellar masses for members of these cores, as implied by \citet{Chiang2017}.
        
\begin{figure}
        \centering
        \includegraphics[width=\columnwidth]{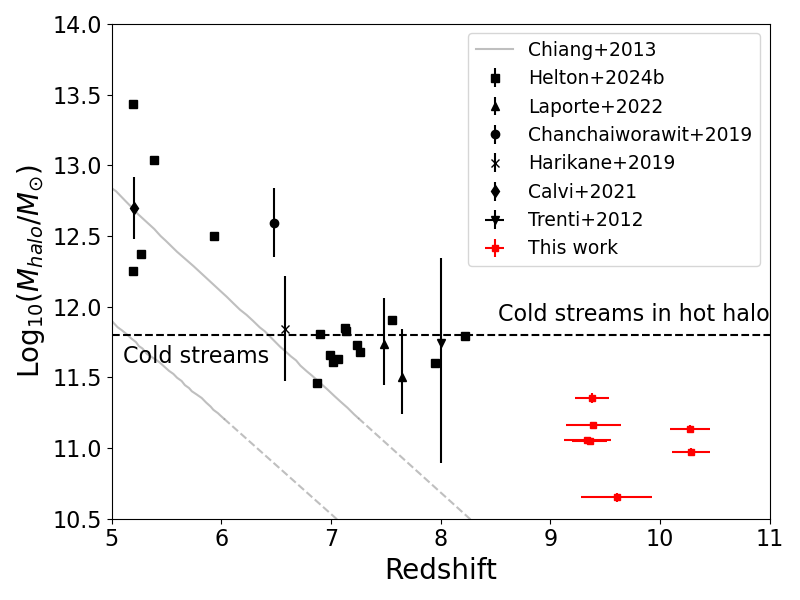}
        \caption{The halo mass of protoclusters as a function of redshift. The area between the grey lines should be the predicted evolution track of a massive cluster from the semi-analytic model by \cite{Chiang2013}, whereas the dashed line is its linear extrapolation. The data points are the observed protoclusters from \citep{Trenti2012, Chanchaiworawit2019, Calvi2021, 2019Harikane, Laporte2022, Helton2024b}. The red stars indicate the seven overdense regions of our work.
        The dashed line illustrates the typical threshold mass for a stable shock in a spherical infall. Below this threshold, the flows are predominantly cold, while above it, a shock-heated medium is present \citep{Dekel2006}.}
        
        \label{fig:HMC}
        \end{figure}

        We convert the stellar mass of each protocluster candidate to the estimated halo mass using the empirical relation from the semi-analytic simulation \cite{Behroozi2019}. However, they only show the relation up to $z = 8$, so we linearly extrapolate the evolution track from $z = 7$ to $8$ and derive the halo-stellar mass ratios at $z \sim 9$.
        Because these galaxies live in subhalos that are still merging into a larger, not fully virialised structure, we estimate the total protocluster halo mass by summing the stellar masses of all member galaxies (from SED fitting) and then applying a halo-stellar mass ratio calibrated at the group/cluster scale, rather than treating each galaxy’s halo individually.
        The extrapolated ratio sits between 0.012 and 0.004, which is also included in the upper- or lower errors in Figure \ref{fig:HMC}.
        
        We also compare the halo mass as a function of the redshift in the previous literature at $z > 5$ \citep{Trenti2012, Chanchaiworawit2019, Calvi2021, 2019Harikane, Laporte2022, Helton2024b}.
        The halo mass inferred from \cite{Laporte2022} is given with (without) the magnification of the foreground galaxy cluster, which becomes $M_{h}/M_{\odot} = 3.34^{+0.59}_{-0.50} \times 10^{11}$ ($3.6^{+13.3}_{-2.8} \times 10^{11}$).
        \cite{Helton2024b} use the simulation result from \texttt{UniverseMachine} to derive the relation of stellar-to-halo mass for their sample ($11.5 < M_{h}/M_{\odot} < 13.4$).
        The halo mass of their most distant protocluster ($z_{\text{spec}} = 7.954$ and $8.222$) ranges $11.5 < M_{h}/M_{\odot} < 12.0$ as a decreasing trend toward higher redshifts.
        The brightest galaxy in the $z = 8.5$ protocluster from BoRG survey has a halo mass of $M_{h}/M_{\odot} = 4-7 \times 10^{11}$ \citep{Trenti2012}.
        
        The total halo mass for each protocluster candidate obtained by combining each galaxy spans from $10^{10.62}$ to $10^{11.39}M_{\odot}$ for our candidates and is shown in Figure \ref{fig:HMC}.
        In this sense, our estimated halo masses are comparable to or exceed the inferred progenitor masses of present-day Coma-like clusters (i.e., $M_{h} > 10^{15} M_{\odot}$ at $z = 0$), suggesting that some of these overdense regions may evolve into similarly massive systems, though a wide range of evolutionary outcomes remains possible.

\section{Conclusions}
\label{Conclusion}
    \begin{enumerate}
        \item We found seven protocluster candidates between $9 \leq z \leq 11$, as galaxies within overdense regions with $\delta$ values exceeding the 95th percentile of the $\delta$ distribution for 366 F115W dropout galaxies. Each region has statistically significant peaks based on Monte Carlo sampling ($\sigma$ > 3).  Because of the large area coverage of the COSMOS-Web, this work presents the largest number of protocluster candidates at $z\gtrsim 9-10$ to date.
        \item  The size of our protocluster candidates ranges $3-4'$ ($\sim$15 cMpc, at $z = 9.3$). These values are comparable to the potential physical sizes of protoclusters by simulations from \citep{Chiang2017} ($\lesssim10$ cMpc).
         However, observations of protocluster candidates at scales of less than 10 cMpc, such as those from \citet{Helton2024b} at $z=8.22$ and \citet{Trenti2012} at $z=8.0$, would mainly target the central galaxies of the protoclusters at $z\geq8$.
        
        \item  We estimate halo masses of protocluster candidates at $z \sim 9-10$ by extrapolating the \citet{Behroozi2019} stellar-to-halo mass relation (assume extra 10\% uncertainty), deriving total halo masses as $10^{10.62-11.39}M_{\odot}$ of our candidates including 1$\sigma$ error; this suggests that all the overdense regions may evolve into Coma-like clusters $M_{h} > 10^{15} M_{\odot}$ at $z = 0$.
    \end{enumerate}

    These overdensities, previously unexplored in the literature, offer a unique opportunity to advance our understanding of the formation and early evolution of protoclusters at $z\sim 9$. 
    All of these findings still require NIRCam spectroscopic follow-up to confirm (or reject) the physical association of the candidate members.
    Shortly, COSMOS-3D (GO 5893; PI Kakiichi) will have NIRCam WFSS grism observations with F444W for the COSMOS field. These grism data would further confirm the protoclusters and high-z candidates spectroscopically, especially the F444W could recover the important H$\beta$ and [OIII]$\lambda$5007 lines for galaxies at $z\sim7-9$.
    If any of the candidates are confirmed spectroscopically, their possible evolution in different environments would examine their role in shaping the large-scale structure of the Universe during the epoch of reionisation.

\begin{acknowledgement}
The authors express their gratitude to the anonymous referee for the very constructive and insightful comments that have significantly improved the quality of this manuscript.
The authors also express their gratitude to the staff in the JWST Helpdesk for their support throughout the process. 
TG acknowledges the support of the National Science and Technology Council of Taiwan through grants 113-2112-M-007-006-, 113-2927-I-007-501-, and 113-2123-M-001-008-.
TH acknowledges the support of the National Science and Technology Council of Taiwan through grants 110-2112-M-005-013-MY3, 110-2112-M-007-034-, and 112-2123-M-001-004-.
SH acknowledges the support of the Australian Research Council (ARC) Centre of Excellence (CoE) for Gravitational Wave Discovery (OzGrav) project numbers CE170100004 and CE230100016, and the ARC CoE for All Sky Astrophysics in 3 Dimensions (ASTRO 3D) project number CE170100013.

This work is based on observations made with the NASA/ ESA/CSA James Webb Space Telescope. The data were obtained from the Mikulski Archive for Space Telescopes at the Space Telescope Science Institute, which is operated by the Association of Universities for Research in Astronomy, Inc., under NASA contract NAS 5-03127 for \textit{JWST}.

This work used high-performance computing facilities operated by the Centre for Informatics and Computation in Astronomy (CICA) at National Tsing Hua University. This equipment was funded by the Ministry of Education of Taiwan, the National Science and Technology Council of Taiwan, and the National Tsing Hua University.
\end{acknowledgement}

\section*{Data Availability Statement}
The COSMOS-Web DR0.5 is publicly available at
\url{https://cosmos.astro.caltech.edu/page/cosmosweb-dr}.
The JADES data used in this study are publicly available through the Mikulski Archive for Space Telescopes (MAST) as part of the High-Level Science Products (HLSP) release:
\url{https://archive.stsci.edu/hlsp/jades} (DOI: 10.17909/8tdj-8n28e).
The JADES DR3 catalogue is available at \url{https://jades-survey.github.io/scientists/data.html}.
All processed data products and analysis scripts used in this work will be made available upon reasonable request to the corresponding author.


\printendnotes

\printbibliography

\end{document}